\documentclass{article}

\usepackage{arxiv}

\usepackage[utf8]{inputenc} 
\usepackage[T1]{fontenc}    
\usepackage{hyperref}       
\usepackage{url}            
\usepackage{booktabs}       
\usepackage{amsfonts}       
\usepackage{nicefrac}       
\usepackage{microtype}      
\usepackage{lipsum}		
\usepackage{graphicx}
\usepackage[numbers]{natbib}
\usepackage{doi}
\usepackage{amsmath}

\title{Suitable Measures for the Potential Operational Utility of AI NWP Rainfall Forecasts Over Africa}

\author{ \href{https://orcid.org/0009-0007-4449-1621}{\includegraphics[scale=0.06]{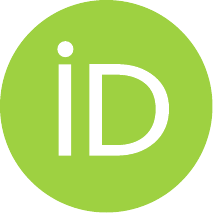}\hspace{1mm}Shruti Nath}\\
	Department of Physics,\\
	University of Oxford, United Kingdom\\
    and\\
	AfriClimate AI, South Africa \\
	\texttt{shruti@africlimate.ai} \\
	\And
	\href{https://orcid.org/0009-0006-2021-7145}{\includegraphics[scale=0.06]{orcid.pdf}\hspace{1mm}Docko Sow} \\
	AfriClimate AI, South Africa\\
    \And
	\href{https://orcid.org/0009-0008-5800-6970}{\includegraphics[scale=0.06]{orcid.pdf}\hspace{1mm}Koomi Toussaint Amoussouvi} \\
	AfriClimate AI, South Africa\\
    \And
	\href{https://orcid.org/0000-0003-3411-2628}{\includegraphics[scale=0.06]{orcid.pdf}\hspace{1mm}Fenwick Cooper} \\
	Department of Physics,\\
	University of Oxford, United Kingdom\\
    \And
	\href{https://orcid.org/0009-0006-5840-7796}{\includegraphics[scale=0.06]{orcid.pdf}\hspace{1mm}Josiah Kiarie Kimani} \\
	AfriClimate AI, South Africa\\
    \And
	\href{https://orcid.org/0009-0009-6105-6458}{\includegraphics[scale=0.06]{orcid.pdf}\hspace{1mm}John Bagiliko} \\
	AfriClimate AI, South Africa\\
    \And
	\href{https://orcid.org/0000-0003-1766-2898}{\includegraphics[scale=0.06]{orcid.pdf}\hspace{1mm}Florian Pappenberger} \\
	ECMWF,\\
    Reading, United Kingdom
    \And
	\href{https://orcid.org/0000-0002-7337-9176}{\includegraphics[scale=0.06]{orcid.pdf}\hspace{1mm}Rendani Mbuvha} \\
	School of Statistics and Actuarial Science,\\
    University of Witwatersrand, South Africa\\
    and\\
    AfriClimate AI, South Africa
}

\renewcommand{\shorttitle}{}

\hypersetup{
pdftitle={A template for the arxiv style},
pdfsubject={q-bio.NC, q-bio.QM},
pdfauthor={David S.~Hippocampus, Elias D.~Striatum},
pdfkeywords={First keyword, Second keyword, More},
}

\begin{document}
\maketitle

\begin{abstract}
Artificial Intelligence (AI)-based weather prediction has advanced rapidly, with models now rivaling physical Numerical Weather Prediction (NWP) systems at a fraction of the computational cost. This is particularly promising for Africa, where rainfall extremes are becoming more frequent and severe, but many forecasting centres lack the infrastructure to run physical NWPs at extended lead times. However, the operational utility of AI NWPs for African rainfall remains uncertain because these models are trained on reanalysis data with known biases and operate at resolutions too coarse to resolve convective rainfall systems. We present a fair, calibrated comparison of GraphCast, GenCast and the Functional Generative Network (FGN) against the physical NWP model IFS. We apply the same postprocessing method, Isotonic Distributional Regression (IDR), to deterministic and probabilistic forecasts, producing calibrated Continuous Ranked Probability Scores (CRPS) against three observational datasets: IMERG, RFEv2 and CHIRPS. Performance is evaluated across seasons, wet and dry regimes, elevation zones and lead times. This framework places all models on an equal footing and estimates their predictive performance under optimal calibration. Across most seasons, all models retain calibrated skill beyond climatology at extended lead times. AI NWPs consistently outperform IFS in wet regions, whereas IFS performs better in dry, high-elevation areas, where its finer grid better resolves orographic controls on rainfall. Spatial analysis shows that AI NWPs achieve a median improvement of approximately 5$\%$ over IFS across observational products and seasons. Despite their architectural differences, deterministic GraphCast achieves calibrated skill comparable to the ensemble-based FGN over their common evaluation period, indicating that probabilistic calibration can narrow apparent performance gaps between AI NWP designs. Nevertheless, FGN provides significantly greater skill at longer lead times, highlighting the value of ensemble forecasting for limiting error growth with increasing prediction horizon.
\end{abstract}


\section{Introduction}
Following years of slow but steady improvements \citep{Bauer2015}, Numerical Weather Prediction (NWP) is in the midst of what might fairly be described as a noisy revolution. Artificial Intelligence (AI) based NWP models such as GraphCast \citep{Lam2023}, GenCast \citep{Price2025} and the Functional Generative Network (FGN) \citep{Alet2025} have progressed rapidly from proof-of-concept demonstrations to systems that rival, and by some verification metrics exceed, the skill of operational physical NWPs such as the Integrated Forecasting System (IFS) \citep{rasp2024weatherbench,Brenowitz2025,Gneiting2026}, all while running at a small fraction of the computational cost. This progress, while genuine, is accompanied by a rapid pace of developments as well as a shift from physics- to data- driven modelling paradigms, which has limited the ability of operational centres to absorb and adopt them. Hence, while promising, the true benefits of applying AI NWPs within real-time operations remain far less apparent. Moreover, with newer AI NWP generations, training strategies and headline skill claims continuing to emerge, rapid, operationally relevant model evaluation struggles to keep pace, and despite certain efforts \citep{rasp2024weatherbench,BenBouallegue2024} remains a critical gap within operational centres.

Due to their computational affordability and potential skill gain, AI NWPs carry particular significance for Africa. Rainfall extremes across the continent are becoming both more frequent and more severe \citep{Bobde2024,coughlan2026}, with notably unusual dynamics driving occurrences even beyond expected return periods \citep{Engel2017,Vondou2026}. At the same time many operational centres over the region lack the infrastructure to run traditional physical NWP systems at a resolution fine enough to resolve locally relevant weather, a constraint that has also historically limited the development of higher-resolution physical models tailored to the continent \citep{Lamptey2024}. AI NWP therefore offers a route to leapfrog the computational component of this infrastructure gap, extending skillful forecasting capability to centres and communities of the region. It does not, however, relieve the dependence on observations: data-driven systems inherit their initial conditions from an observation-constrained analysis, and there is growing evidence that their skill degrades at least as steeply as that of physical models when observational coverage is reduced. Compute and observations are therefore complements rather than substitutes in this context. To this end, the potential of AI NWPs comes with caveats specific to the African context. Most AI NWP models are trained on reanalysis datasets such as ERA5, which has known biases in representing rainfall over Africa, stemming from inadequate observational coverage \citep{dinku2019,liu2024}. Current global AI NWP models also operate at a resolution of around 25~km, which is too coarse to represent the deep convective systems that drive much of Africa's rainfall \citep{Fink2017,Gudoshava2024Advances}. The actual operational utility of AI NWPs within the African context, and particularly for rainfall, therefore cannot be inferred from global evaluation alone and requires downstream post-processing for operationally relevant evaluation. It is well known that statistical post-processing of NWP output can improve forecast skill and reliability, particularly over tropical Africa \citep{Vannitsem2021,Vogel2018,Vogel2020}. Furthermore, comparisons of physics-based against classical Machine Learning (ML) to AI approaches for forecasting rainfall and rainfall occurrence over Northern Tropical Africa have shown that ML/AI methods can match or exceed the performance of classical physics-based models \citep{Vogel2021,PhysicsBasedvsDataDriven24HourProbabilisticForecastsofPrecipitationforNorthernTropicalAfrica}. To this extent, several studies highlight that combining NWP based forecasts with probabilistic post-processing methods can provide more impartial comparisons across physical- and AI- based approaches \citep{Cooper2025,Gneiting2026,nath2026predicting}. 

Here, we recognise that a fair assessment of the potential operational utility of NWP forecasts should imitate the quality an actual forecast would have upon reaching an operational forecaster's desk. This requires that any AI or physical NWP output first be post-processed to match the resolution most relevant to a given national meteorological centre, as well as corrected towards suitable, local observational products. Such post-processing should also yield calibrated, probabilistic outputs, allowing for a fairer comparison between AI NWPs, whose exact nature of ensembling is architecture driven and different in nature as compared to physical NWPs, whose ensembling is mostly initial condition driven \citep{Brenowitz2025,Gneiting2026}. We employ an Isotonic Distributional Regression (IDR) post-processing \citep{Henzi2021,Walz2024} to obtain calibrated Continuous Ranked Probability Scores (CRPS) for each AI NWP rainfall forecasts over Africa, considering GraphCast, GenCast and FGN alongside the physical NWP, IFS. against three independent observational products, IMERG, RFEv2 and CHIRPS, evaluated jointly across seasons, wet and dry regimes and elevation bands. Considering multiple observational products rather than a single reference dataset allows us to distinguish genuine differences in forecast skill from artefacts specific to any one verification product, an important distinction given the known disagreement among African rainfall datasets described above \citep{Ageet2022,mekonnen2023,mbuvha2026}. 

The rest of this paper is structured as follows. Section~\ref{sec2} introduces the models (Section~\ref{sec2a}), observational products (Section~\ref{sec2b}) and verification methodology (Section~\ref{sec2c}) in detail. Section~\ref{sec3} then provides Results divided into overall performance across seasons and elevations (Section~\ref{sec3a}) and detailed spatial analysis of the relative differences in skill between AI NWPs and IFS (Section~\ref{sec3b}). We then proceed to Conclusions and Discussions in Section~\ref{sec4} and provide some Outlook in Section~\ref{sec5}.

\section{Methods}\label{sec2}

\subsection{Numerical Weather Prediction (NWP) models}\label{sec2a}

The physical and AI NWP models used are described in the following sub-sections. It should be noted that for all models, evaluation is done on the native resolution of each rainfall observational product (Section~\ref{sec2b}). Where necessary, rainfall forecasts are therefore re-gridded using bilinear interpolation, to match the native resolution  prior to verification. Years available for each NWP product is further summarised in Table \ref{tab1}, and we mainly consider years after 2020 which are largely out of the training data of the AI NWP models although have some overlap with the 2016-2022 period used in operational fine-tuning. For this study, we nevertheless favour having a larger sample of years for evaluation and allow for some data leakage to that extent.

\begin{table}[!ht]
\caption{Years available for each model dataset, we only consider years that are within the AI models test set (2020 and above)}
\centering
\begin{tabular}{l c c c c c c c}
\hline
Model & 2020 & 2021 & 2022 & 2023 & 2024 & 2025\\
\hline
IFS-ENS & \checkmark & \checkmark & \checkmark & x & \checkmark & \checkmark  \\
GraphCast & \checkmark & \checkmark & \checkmark & \checkmark & \checkmark & \checkmark \\
GenCast & \checkmark & \checkmark & \checkmark & \checkmark & \checkmark & \checkmark \\
FGN & x & x &  \checkmark & \checkmark & \checkmark & \checkmark 
\\
\hline
\end{tabular}
\label{tab1}
\end{table}

\subsubsection{Physical NWP benchmark}\label{sec2a1}
As our physical NWP benchmark, we consider ECMWF's Integrated Forecasting System ensemble prediction system (IFS), which comprises 50 perturbed members whose spread represents forecast uncertainty arising from both initial-condition perturbations and a representation of model uncertainty through stochastic parametrisation \citep{Leutbecher2008}. The model itself operates on a cubic-octahedral reduced Gaussian grid and following the Cycle 48r1 upgrade of 27 June 2023, had its resolution increased from TCo639 ($\sim$18~km) to TCo1279 ($\sim$9~km), matching that of the deterministic, high-resolution IFS version. The required IFS total precipitation fields were obtained primarily from the WeatherBench2 archive \citep{rasp2024weatherbench}, with remaining fields and periods retrieved directly from ECMWF's Meteorological Archival and Retrieval System (MARS). Beyond this resolution change, the IFS underwent further cycle upgrades within the evaluation period, including changes to moist physics that affect precipitation. The IFS is therefore not a stationary benchmark across 2020-2025, and small differences between systems should be interpreted with that in mind. Looking forward, hybrid configurations in which a physical model is spectrally nudged towards a machine-learned forecast at large scales are under active development \citep{Husain2024,Polichtchouk2026}; where such configurations become operational, the distinction between physical and AI benchmarks will blur, and future studies of this kind will need to specify the model cycle rather than the model name.

\subsubsection{Deterministic AI NWPs}\label{sec2a2}

GraphCast \citep{Lam2023} is a deterministic, autoregressive machine-learning weather model based on a graph neural network (GNN) with an encode--processor--decode architecture: the input state on a $0.25^\circ$ grid is encoded onto an icosahedral mesh, processed via learned message passing, and decoded back to the grid. The operational configuration has 37 pressure levels, 16 processor layers and $\sim$$3.67\times10^{7}$ parameters, and produces forecasts autoregressively at a 6-hour timestep from the two most recent states.

GraphCast is trained on ERA5 (1979--2017) by minimising a latitude- and level-weighted mean-squared-error (MSE) loss,
\begin{equation}
\mathcal{L}_{\mathrm{MSE}} = \frac{1}{|D_{\mathrm{batch}}|}\sum_{d_0\in D_{\mathrm{batch}}}\frac{1}{T_{\mathrm{train}}}\sum_{\tau=1}^{T_{\mathrm{train}}}\frac{1}{|G_{0.25^\circ}|}\sum_{i\in G_{0.25^\circ}}\sum_{j\in J} s_j\, w_j\, a_i\,\big(\hat{x}^{d_0+\tau}_{i,j}-x^{d_0+\tau}_{i,j}\big)^2 ,
\label{eq:graphcast-loss}
\end{equation}
with the autoregressive rollout length $T_{\mathrm{train}}$ increased during training from 1 to 12 steps, and for this study we consider the GraphCast version further fine-tuned on HRES-fc0 for operational use. 

\subsubsection{Probabilistic AI NWPs}\label{secn2a3}

\paragraph{GenCast.}
GenCast \citep{Price2025} is a conditional diffusion model that generates probabilistic ensembles by iteratively denoising a candidate state, conditioned on the two previous states, and sampling independent stochastic trajectories:
\begin{equation}
p\big(X^{1:T}\,|\,X^{0},X^{-1}\big)=\prod_{t=0}^{T-1}p\big(X^{t+1}\,|\,X^{t},X^{t-1}\big).
\label{eq:gencast-factorisation}
\end{equation}
It reuses the GraphCast GNN deterministic backbone (latent dimension 512, 16 layers, $\sim$$5.7\times10^{7}$ parameters), conditioned on the diffusion noise level via conditional layer normalisation, and operates at $0.25^\circ$ resolution with a 12-hour timestep. As we consider the operational product for this study, we use the GenCast version that has since been fine-tuned on HRES-fc0.

\paragraph{FGN.}
FGN, the Functional Generative Network \citep{Alet2025}, similarly targets the trajectory distribution
\begin{equation}
p\big(X^{1:T}\,|\,X^{0},X^{-1}\big)=\int p(\epsilon)\prod_{t=1}^{T}p\big(X^{t}\,|\,X^{t-2:t-1},\epsilon\big)\,\mathrm{d}\epsilon,
\label{eq:fgn-factorisation}
\end{equation}
at a 6-hour timestep and $0.25^\circ$ resolution, but generates ensembles via functional (parameter-space) perturbations rather than diffusion. Epistemic uncertainty is captured by a deep ensemble of $J=4$ independently trained models; aleatoric uncertainty is captured by sampling network parameters via the reparameterisation trick,
\begin{equation}
\theta = \theta^\ast + \Delta\cdot\epsilon,
\label{eq:fgn-reparam}
\end{equation}
where a single 32-dimensional global noise vector $\epsilon\sim\mathcal{N}(0,I_{32})$, scaled by a learned matrix $\Delta$ that maps it into parameter space, is fed into all conditional layer-norm layers, with parameters shared across the mesh and grid. This low-dimensional, globally shared perturbation encourages coherent spatial variability. FGN is trained on the per-location marginal, ``fair" Continuous Ranked Probability Score (CRPS), with a correction applied to de-bias for number of ensemble members \citep{Ferro2014,Zamo2018}. FGN is larger than GenCast ($\sim$$1.8\times10^{8}$ parameters/seed, latent dimension 768, 24 processor layers) and is trained in stages of increasing resolution on ERA5 before autoregressive fine-tuning on HRES-fc0.

\subsection{Observational Data}\label{sec2b}

Precipitation over Africa is notoriously difficult to observe, owing to a sparse and declining rain-gauge network \citep{dinku2018,Su2026}, and no single dataset can be treated as unambiguous truth. We therefore verify against three independent, widely used satellite--gauge blended precipitation products, each with different input streams and bias characteristics, and treat agreement (or disagreement) among them as an indication of observational uncertainty \citep{Pappenberger2009}. These products were chosen on the basis of having the best, most consistent performance across the continent following a systematic review of previous studies \citep{mbuvha2026}. All verification statistics in Section~\ref{sec2c} are computed independently against each of the three products, and results are compared across products rather than pooled, so that conclusions are not an artefact of any single observational dataset.

\subsubsection{Integrated Multi-satellitE Retrievals for GPM (IMERG)}\label{sec2b1}

IMERG is the NASA/JAXA Global Precipitation Measurement (GPM) mission's Level-3 multi-satellite precipitation product \citep{Huffman2020}, estimating global surface precipitation at $0.1^\circ$ spatial resolution every 30 minutes from June 2000 to the present. Passive-microwave estimates from the GPM constellation are computed with the Goddard Profiling Algorithm (GPROF), gridded, intercalibrated to the GPM Combined Radar--Radiometer (CORRA) product, morphed and blended with geostationary infrared estimates. IMERG is disseminated in three runs of increasing latency and accuracy: the Early Run (4-hour latency), the Late Run (14-hour latency), and the Final Run (3.5-month latency). We use the Final Run (V07), in which the satellite estimates are adjusted to sum to a monthly satellite--gauge combination that incorporates the Global Precipitation Climatology Centre (GPCC) monthly gauge analysis, this is the most accurate, most heavily corrected IMERG product and is the standard research baseline. IMERG's fine resolution and full tropical coverage make it particularly valuable over data-sparse Africa, though the documentation notes reduced skill over complex terrain, coastal zones and frozen surfaces, and possible artefacts from the evolving satellite constellation.

\subsubsection{Climate Hazards Group InfraRed Precipitation with Station data (CHIRPS)}\label{2b2}

CHIRPS \citep{Funk2015} is a quasi-global ($50^\circ$S--$50^\circ$N), $0.05^\circ$-resolution, 1981-to-present rainfall dataset produced by the Climate Hazards Center at the University of California, Santa Barbara, and is the most widely used product for African drought and rainfall monitoring. Its algorithm blends three ingredients: a high-resolution monthly precipitation climatology (CHPclim); $0.05^\circ$ satellite-only precipitation estimates derived from thermal-infrared cold-cloud-duration (CCD) observations (the CHIRP product); and in-situ station observations, merged via a spatial-correlation-weighted interpolation. CHIRPS is available at daily, pentadal and monthly aggregations (a preliminary product at about 2-day latency and a final product at roughly 3-week latency). It was explicitly designed to support seasonal drought monitoring and hydrologic trend analysis in regions such as the Greater Horn of Africa, and validation over eastern Africa \citep{dinku2018} has found it to perform well; its accuracy nonetheless degrades in station-sparse rural regions and it is land-only.

\subsubsection{Rainfall Estimate (RFE) version 2}\label{sec2b3}

RFEv2 is the operational daily rainfall product of the NOAA Climate Prediction Center (CPC), developed to support the U.S.\ Agency for International Development's Famine Early Warning Systems Network (FEWS NET) over Africa. Implemented operationally at CPC in 2001 on the basis of the methods of Xie and Arkin \citep{Xie1997}, RFEv2 provides daily (0600--0600~UTC) estimates at $0.1^\circ$ resolution over the African continent by combining four inputs: geostationary Meteosat infrared imagery, polar-orbiting passive-microwave estimates from SSM/I and AMSU-B, and quality-controlled Global Telecommunication System (GTS) rain-gauge reports. Its methodology is documented by Novella and Thiaw \citep{Novella2013}; a known limitation, noted by those authors is that low GTS station reporting rates degrade estimates in some regions.

\subsection{Verification}\label{sec2c}

\subsubsection{Potential Continuous Ranked Probability Score}\label{sec2c1}

The primary verification metric is the Continuous Ranked Probability Score (CRPS), which is a strictly proper scoring rule \citep{Gneiting2007}. For a forecast cumulative distribution function $F$ and a scalar observation $y$ the CRPS is given as,
\begin{equation}
\mathrm{CRPS}(F,y)=\int_{-\infty}^{\infty}\big(F(z)-\mathbb{1}[z\ge y]\big)^2\,dz,
\label{eq:crps-def}
\end{equation}
with lower values indicating better forecasts, and units matching those of the forecast variable (mm~h$^{-1}$ for precipitation rate in this study). 

For an operational forecaster, a forecast's skill can be decomposed into two components: miscalibration error, which could in part be ameliorated through statistical post-processing, and the forecast's ability to discriminate an event's likelihood beyond a climatological guess. Understanding this decomposition is integral to correctly interpreting a forecast's skill and assessing its operational utility. Using the simple base case of Murphy's Brier skill score decomposition \citep{murphy1973}, a forecast whose reliability (i.e., miscalibration) error is equal to its resolution (i.e., discrimination) would have its overall skill dominated by climatological uncertainty, from which one may argue that it is ``useless" beyond climatology \citep{Mason2004, weisheimer2014}. Methods of translating Murphy's decomposition to a broader range of proper scoring rules have been explored \citep{Hersbach2000,dawid2004,kull2015,Siegert2017}. In this study we consider the score decomposition introduced by Dimitriadis \citep{Dimitriadis2021}, which is simple and interpretable,
\begin{equation}
\mathrm{CRPS}=\underbrace{(\mathrm{CRPS}-\mathrm{CRPS}_{\mathrm{cal}})}_{miscalibration}-\underbrace{(\mathrm{CRPS}_{\mathrm{ref}}-\mathrm{CRPS}_{\mathrm{cal}})}_{discrimination}+\underbrace{\mathrm{CRPS}_{\mathrm{ref}}}_{uncertainty}
\label{eq:crps-decomp}
\end{equation}
Where $\mathrm{CRPS}_{\mathrm{ref}}$ is the skill of a reference forecast \citep{Pappenberger2015b}, which for this study is considered as the climatological skill $\mathrm{CRPS}_{\mathrm{clim}}$, calculated for each observational product separately, with climatology constructed as according to \citep{Walz2021}.  $\mathrm{CRPS}_{\mathrm{cal}}$, is the score the system would attain if it were optimally calibrated to achieve perfect reliability. For this study $\mathrm{CRPS}_{\mathrm{cal}}$ is used as a proxy of the skill under real-time use for operational probabilistic forecasting subject to statistical post-processing \citep{Gneiting2026}. 

The distinction of $\mathrm{CRPS}$ from $\mathrm{CRPS}_{\mathrm{cal}}$ is central to a fair comparison between AI and physical NWP systems given their systematically different, and partly unknown, calibration properties \citep{Brenowitz2025}. FGN is trained directly on a CRPS objective, GenCast implicitly encodes calibration through its diffusion sampling process, and IFS derives its spread from a physically-motivated but distinct perturbation scheme, whereas GraphCast is the only model considered that is purely deterministic. Rather than assume any one system is well calibrated, we estimate $\mathrm{CRPS}_{\mathrm{cal}}$ directly for every model; GraphCast, GenCast, FGN and IFS alike. We follow an isotonicity-based decomposition, which uses an IDR \citep{arnold2023decompositionsmeancontinuousranked,Gneiting2007} to recalibrate model outputs as further described in Section~\ref{sec2c2}, and report on the obtained calibrated performance, $\mathrm{CRPS}_{\mathrm{cal}}$, throughout the results, with differences between $\mathrm{CRPS}$ and $\mathrm{CRPS}_{\mathrm{cal}}$ provided in Figures \ref{fig:s1} and \ref{fig:s2}. The isotonicity-based decomposition ensures non-negativity in the obtained miscalibration and discrimination terms, however, we note that this only holds for in-sample recalibration; under the blocked out-of-sample scheme adopted here (Section~\ref{sec2c2}) it is no longer guaranteed, and small negative miscalibration values may arise. 

Two summary quantities built on $\mathrm{CRPS}_{\mathrm{cal}}$ are used. First, to assess the potential discrimination relative to climatology we report,
\begin{equation}
\Delta\mathrm{CRPS}_{\mathrm{pot}} = (\mathrm{CRPS}_{\mathrm{cal}}-\mathrm{CRPS}_{\mathrm{clim}}),
\label{eq:crps-diff-clim}
\end{equation}
in the native units of the observational product (mm~h$^{-1}$). 
$\Delta\mathrm{CRPS}_{\mathrm{pot}}$ is equal to the negative of the discrimination term from Equation~\ref{eq:crps-decomp} and is negatively oriented, such that $\Delta\mathrm{CRPS}_{\mathrm{pot}}>0$ indicates performance worse than climatology. Second, to compare AI models directly against the physical benchmark on a normalised, dimensionless scale (Section~\ref{sec2c3}), we define the potential CRPS skill score (CRPSS) with respect to IFS,
\begin{equation}
\mathrm{CRPSS} = 1-\frac{\mathrm{CRPS}_{\mathrm{cal}}(\text{model})}{\mathrm{CRPS}_{\mathrm{cal}}(\text{IFS})},
\label{eq:crpss-ifs}
\end{equation}
where $\mathrm{CRPSS}>0$ indicates improved resolution relative to IFS and $\mathrm{CRPSS}<0$ indicates degraded resolution.

Statistical significance of $\mathrm{CRPSS}$ at each grid cell is assessed with the Diebold--Mariano (DM) test \citep{DieboldMariano1995}, applied to the paired time series of per-forecast $\mathrm{CRPS}_{\mathrm{cal}}$ loss differentials between a given AI NWP and IFS. The DM test is conducted for each prediction lead-time independently and to account for serial auto-correlation we apply the small-sample correction of Harvey et al. \citep{Harvey1997}. The null hypothesis of the DM test is that the two forecasting systems have equal predictive accuracy (zero expected loss differential). We reject the null hypothesis, and treat the local skill differential as significant, where the resulting $p$-value falls below $0.1$. Grid cells for which the null hypothesis is not rejected are flagged (i.e., dotted in spatial maps) to indicate that the sign of the local $\mathrm{CRPSS}$ could plausibly be due to sampling variability rather than a genuine difference in forecast skill.

\subsubsection{Isotonic Distributional Regression (IDR) calibration and Blocked Cross-Validation}\label{sec2c2}

Because GraphCast does not natively produce a predictive distribution, and because a fair $\mathrm{CRPS}_{\mathrm{cal}}$ comparison (Section~\ref{sec2c1}) requires every system to be represented as a calibrated distribution, we convert each model's raw output into a probabilistic forecast using isotonic distributional regression (IDR), implemented via the Easy Uncertainty Quantification procedure \citep{Henzi2021,Walz2024,Gneiting2026}. For a deterministic (or ensemble-summary) forecast value $x$, IDR estimates the conditional Cumulative Distribution Function (CDF) of the observation given the forecast, $\hat F(\cdot \mid x)$, non-parametrically, subject only to the constraint that the estimated conditional distribution is stochastically monotonically non-decreasing in $x$ (i.e.\ a higher forecast value implies a stochastically larger predicted outcome). This isotonicity constraint is the sole structural assumption imposed with no underlying forecast distribution (e.g.\ Gaussian or Gamma). IDR is fitted separately for each model, lead time, and observational product. To ensure enough representative samples for building the predicted CDF, a separate IDR is fitted per Köppen--Geiger climate zone \citep{Beck2018}, pooling together all grid points within each zone and their available timesteps. This follows a space-for-time approach given the limited number of out-of-sample test years available across AI NWPs. Given that Köppen--Geiger climate zones are constructed from temperature and precipitation climatology, we assume that pooling their zonal grid-points is representative of similar underlying rainfall distributions. We note that this assumption is most likely to be strained in regions of sharp climatological gradient such as the Ethiopian Highlands and the Rift Valley rainfall, which is why station-based scores such as SEEPS \citep{Rodwell2010} are normalised against each station's own climatology. 

The resulting calibrated distribution $\hat F$ is used to calculate $\mathrm{CRPS}_{\mathrm{cal}}$ of Section~\ref{sec2c1} and this calibration step is applied uniformly to all four systems considered. For IFS, GenCast and FGN (whose raw forecasts are already ensembles), IDR is applied to the ensemble mean, which was found to be more effective, and computationally cheaper, than using the full ensemble. According to previous studies \cite{Brenowitz2025, Gneiting2026}, $\mathrm{CRPS}_{\mathrm{cal}}$ is likely an optimistic measure of forecast skill at short lead times, since it could rely on in-sample information not available operationally, but a pessimistic measure at longer lead times, when an operational ensemble can draw on additional information (e.g., evolving weather regimes) beyond what is captured in the training data. For this study, we emphasise the use of the test period 2020-2025 which is largely out of training for the AI NWP models. To further remove any artefacts arising from in-sample IDR performance, we adopt a blocked evaluation approach in which multiple IDRs are trained, each with a different block of two consecutive years (e.g., [2020, 2021], [2022, 2023], and [2024, 2025]) held out. The held-out years are then used to compute out-of-sample estimates of CRPS$_{\mathrm{cal}}$, so as to obtain a robust measure of out-of-sample performance across multiple years.

\subsubsection{Decomposition across elevation and areas receiving significant rainfall}\label{sec2c3}

Because forecast skill over Africa is strongly modulated by both terrain and rainfall regime, we stratify the verification spatially and temporally rather than reporting only continental, all-season averages, which are dominated by a small number of high-rainfall regions and can mask regional failures.

\paragraph{Seasonal stratification.}
We partition the year into three four-month seasons that together span the annual cycle of the tropical rain belt over Africa: February--March--April--May (FMAM), June--July--August--September (JJAS), and October--November--December--January (ONDJ). Potential CRPS and CRPSS$_{\mathrm{IFS}}$ (Section~\ref{sec2c1}) are computed separately within each season, at lead times of 1, 3, 5, 7 and 9~days, for each of the three observational products in turn.

\begin{figure}[!ht]
 \centering
\includegraphics[width=0.7\textwidth]{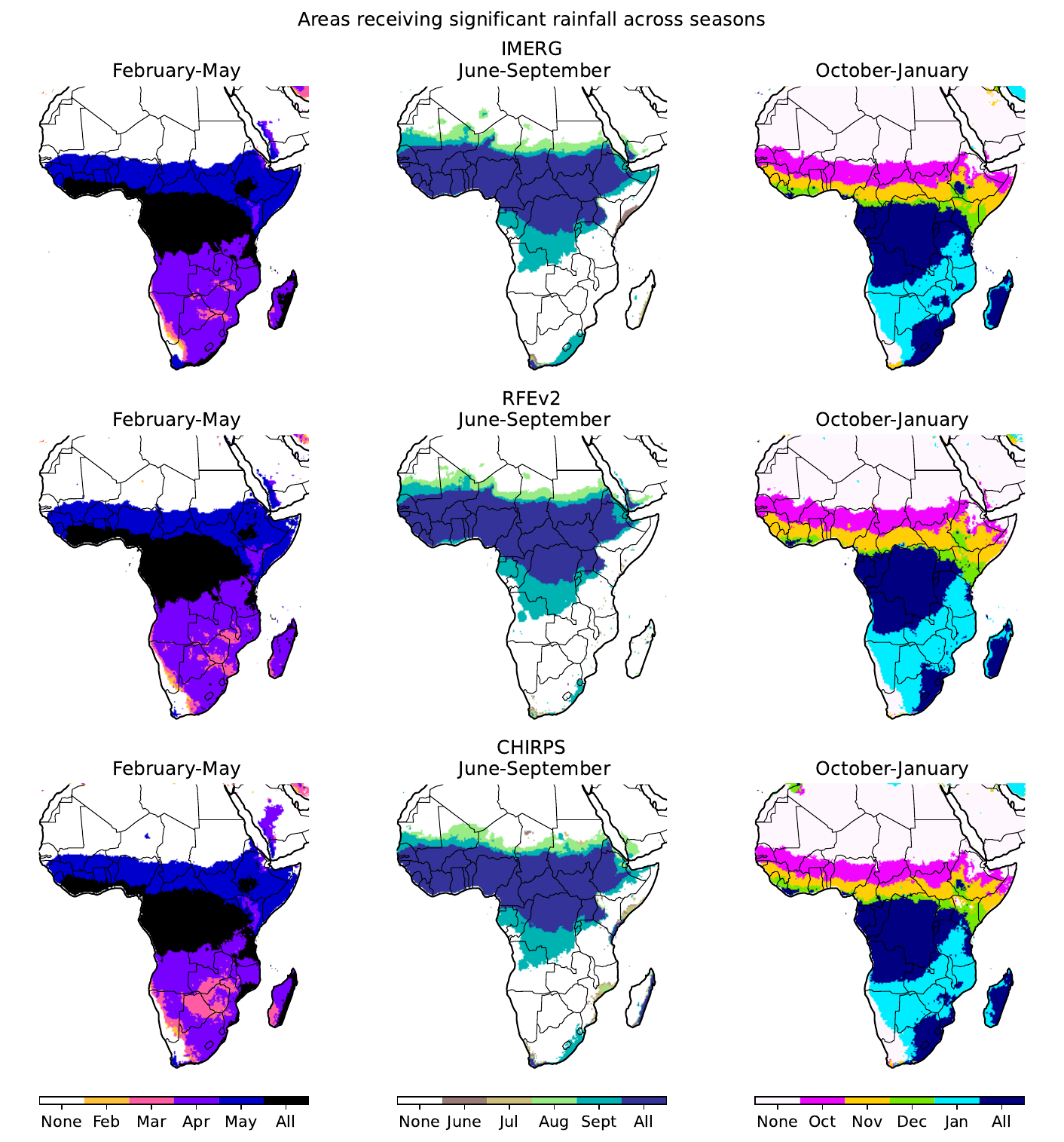}
 \caption{Areas receiving significant rainfall across seasons of February March April and May (FMAM, left); June, July, August and September (JJAS, middle); and October, November, December and January (ONDJ, right). For each season we consider the breakdown of months receiving significant rainfall, defined as on average 1 mm/day or more for a given month. This is calculated for satellite products of IMERG (top), RFEv2 (middle) and CHIRPS (bottom).}
\label{fig:fig1}
\end{figure}

\paragraph{Significant-rainfall-area masking.}
Within each season, grid cells are further classified as receiving ``significant rainfall'' for a given calendar month if the climatological mean precipitation rate for that month, computed from the observational record, is on average 1~mm~day$^{-1}$ or more; grid cells falling below this threshold in a given month are classified as ``dry'' for that month. Because the three observational products (IMERG, RFEv2, CHIRPS) can disagree on which grid cells and months meet this threshold, the mask is computed independently from each product and applied only to the verification performed against that same product, rather than using a single, product-independent mask. Figure~\ref{fig:fig1} shows the resulting significant-rainfall areas for each of the three four-month seasons and each observational product, together with the breakdown by constituent month; the extent and boundaries of the significant-rainfall areas are broadly consistent across IMERG, RFEv2 and CHIRPS, tracing the seasonal migration of the West African monsoon, the Congo Basin convergence zone, and the bimodal long- and short-rains regions of East Africa, though the products disagree in detail, particularly along the drier margins of each wet region. Verification statistics are reported separately for ``all'' grid cells, ``wet'' (significant-rainfall) grid cells only, and ``dry'' grid cells only, so that forecast skill over regions of genuine hydrological and agricultural relevance can be distinguished from skill over predominantly dry areas, where low absolute CRPS values can otherwise give a misleadingly favourable impression of overall performance.

\paragraph{Elevation stratification.}
Coarse-resolution ($0.25^\circ$) AI models, and GraphCast in particular, are known to struggle to resolve orographic precipitation over mountainous and topographically complex terrain. We therefore additionally partition grid points into four elevation bands as depicted in Figure~\ref{fig:fig1}, using a static elevation field consistent with that used by the AI models' native grid: $<500$~m, $500$--$1000$~m, $1000$--$2000$~m, and $>2000$~m. Potential CRPS relative to climatology (equation~\ref{eq:crps-diff-clim}) is computed within each elevation band, restricted to grid cells receiving significant rainfall as defined above, at lead times of 1, 3, 5, 7 and 9~days, for each observational product. This stratification isolates the degree to which AI-model skill deficits are concentrated over high terrain, and whether the probabilistic AI systems (GenCast, FGN) mitigate the deterministic GraphCast's orographic limitations, complementing the seasonal and rainfall-area stratifications above.

\begin{figure}[!ht]
 \centering
\includegraphics[width=0.5\textwidth]{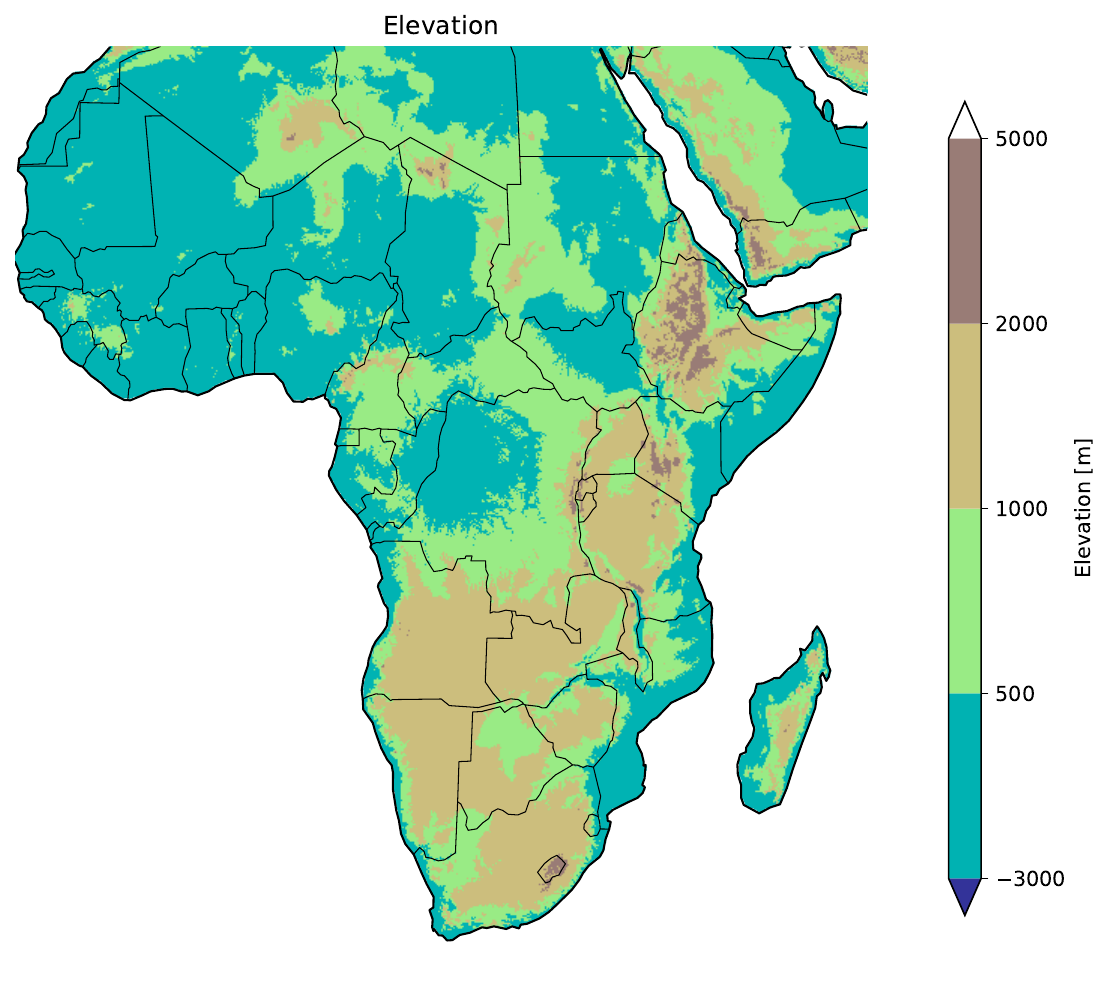}
 \caption{Elevation map over Africa binned according to the stratification of $<500$~m, $500$--$1000$~m, $1000$--$2000$~m, and $>2000$~m used for verification in this study.}
\label{fig:fig2}
\end{figure}

\section{Results}\label{sec3}

We present key findings from benchmarking and evaluation carried out using CRPS$_{\mathrm{cal}}$ as described in Section \ref{sec2c1}. As a first step, we consider the differences between CRPS$_{\mathrm{cal}}$ values calculated for physical and AI NWP models relative to climatology ($\Delta$CRPS$_{\mathrm{pot}}$, Equation \ref{eq:crps-diff-clim}) as an estimate of forecast resolution, and results are shown in Section~\ref{sec3a}.  We then provide detailed comparison of the relative improvement in CRPS$_{\mathrm{cal}}$ with respect to that of the physical NWP benchmark, IFS. The potential CRPS skill score, CRPSS (Equation \ref{eq:crpss-ifs}), is used and results discussed under Section~\ref{sec3b}.

\subsection{Overall performance across seasons and elevations}\label{sec3a}

Beyond continental and annual headline scores, this study aims to provide a granular evaluation of forecast utility and operational relevance. As described in Section~\ref{sec2c3}, we therefore analyse forecast resolution, estimated by $\Delta$CRPS$_{\mathrm{pot}}$, across seasons, elevations, and areas receiving significant rainfall. We estimate uncertainty of $\Delta$CRPS$_{\mathrm{pot}}$ values by taking 100 bootstrap resamples across years and months available, thus considering inter-annual and intra-seasonal variability. For an operational forecaster, this offers nuanced evidence into whether a forecast adds value across the diverse climatic regimes spanning their country. 

\begin{figure}[!ht]
 \centering
\includegraphics[width=0.9\textwidth]{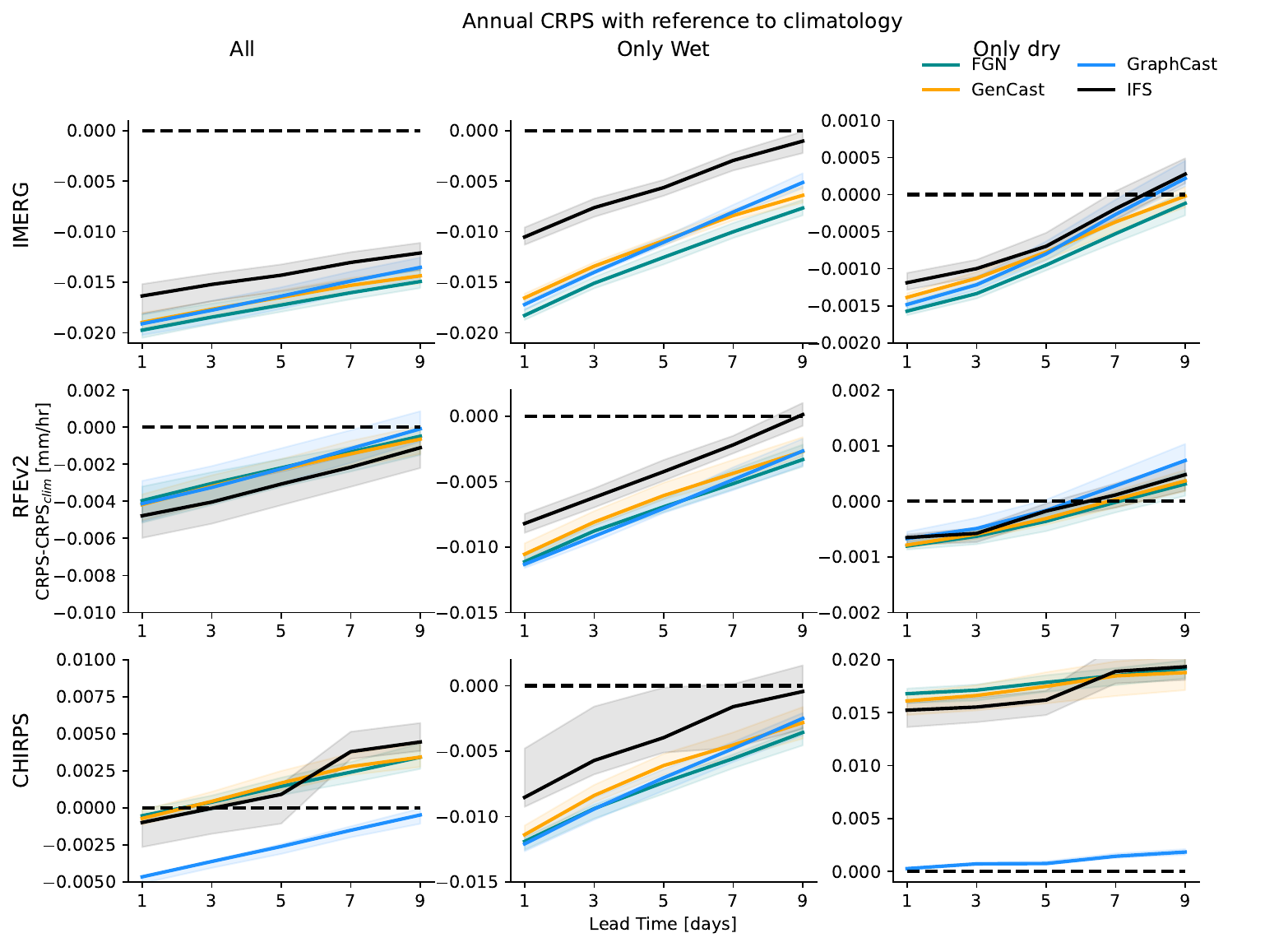}
 \caption{Annual difference in potential Continuous Ranked Probability Score (CRPS) with respect to climatology plotted against lead time and calculated for IMERG (top), RFEv2 (middle) and CHIRPS (bottom). We consider potential CRPS values as averaged across all grid-cells in the domain (All, left), areas only receiving significant rainfall for each month (Wet, middle) and vice versa (Dry, right). A more negative difference indicates lower CRPS with  respect to climatology and better forecast resolution. Uncertainty bands are provided by bootstrap resampling of monthly values across all years considered for each model, therefore representing inter-annual and intra-seasonal uncertainty.}
\label{fig:fig3}
\end{figure}

Figure~\ref{fig:fig3} decomposes $\Delta$CRPS$_{\mathrm{pot}}$ into areas receiving significant rainfall (labelled ``Only Wet") and areas not receiving significant rainfall (labelled ``Only Dry"), where a more negative value indicates better resolution. Results for all areas (All) are given as reference and reveal notable differences across the observational products considered: IMERG shows the AI NWPs having better resolution, beyond range of uncertainty bands, than IFS; RFEv2 shows no significant difference; and CHIRPS shows all models other than GraphCast performing worse than even climatology (i.e., positive $\Delta$CRPS$_{\mathrm{pot}}$). Apportioning these results into ``Only Wet" and ``Only Dry" areas provides useful insights into where these discrepancies could arise from.

``Only Wet" areas show a consistent improvement in resolution across all AI NWP models as compared to IFS, with IMERG showing a steady improvement of degree 0.007, while RFEv2 and CHIRPS show roughly half that magnitude (0.0035). It should be noted, however, that IFS's own improvement over climatology is roughly half as large for RFEv2 and CHIRPS as for IMERG. The smaller gap in performance between AI NWPs and IFS observed for these two products may therefore reflect an observational product-dependent ceiling on achievable improvement over climatology, rather than a smaller benefit of the AI NWPs relative to IFS. For both IMERG and RFEv2, ``Only Dry" areas show negligible differences in skill across all models. CHIRPS, however, shows all models performing worse than climatology, with GraphCast showing the smallest degradation (around 0.001) while the other models show degradations of 0.015–0.02. This suggests that the performance discrepancy seen for CHIRPS under ``All" areas can be predominantly attributed to dry area performances. Such a discrepancy could arise from GraphCast being the only deterministic model amongst those considered. To this extent, instead of hedging across an ensemble of possible outcomes, its training objective favours sharper distinction between wet vs. dry occurrence as compared to its ensemble-based counterparts. This could compound with the known artefacts of CHIRPS itself, where due to the multiplicative correction used, areas with a climatological mean rainfall of 0 are assigned values of 0 \citep{dinku2018}. As a result, by committing to predicting ``no rain", GraphCast better matches CHIRPS internal biases and the better skill compared to other NWPs may therefore be more so artificially induced. We note that a model-side contribution cannot be excluded: both physical and data-driven models are known to over-produce very light precipitation \citep{Stephens2010}, a tendency that is most conspicuous over arid regions and would degrade dry-area scores against any reference. Distinguishing the observational artefact from the model behaviour would require verification against station data in these regions.

\begin{figure}[!ht]
 \centering
\includegraphics[width=0.9\textwidth]{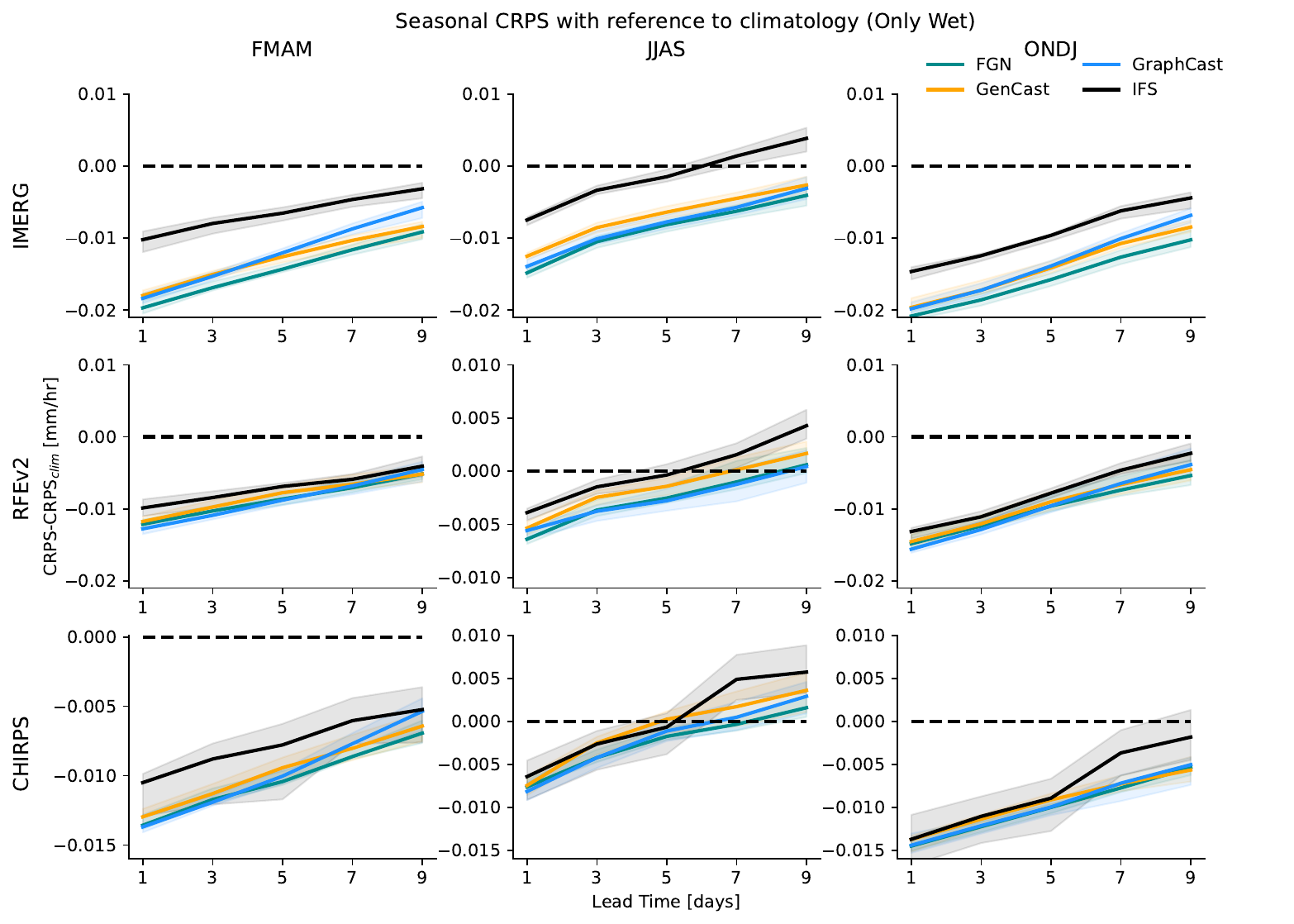}
 \caption{Same as Figure \ref{fig:fig3}, but showing seasonal performance when only considering areas receiving significant rainfall for each month instead. Seasons of February March April and May (FMAM, left); June, July, August and September (JJAS, middle); and October, November, December and January (ONDJ, right) are shown.}
\label{fig:fig4}
\end{figure}

Figure \ref{fig:fig4} decomposes ``Only Wet" area performance into seasons of February March April and May (FMAM), June, July, August and September (JJAS), and October, November, December and January (ONDJ). Consistent with Figure \ref{fig:fig3}, IMERG shows the largest improvement across observational products from AI NWPs as compared to IFS. In contrast to Figure \ref{fig:fig3} however, the performance differential does not always hold constant with lead time. For example, the gap between AI NWPs and IFS narrows with lead time across all observational products for FMAM and similarly for IMERG in ONDJ, while for CHIRPS the gap widens for JJAS and ONDJ. Interestingly, for all observational products and models, JJAS displays the worst skill, with IFS losing resolution beyond climatology at 5-7 day lead times, while AI NWPs lose resolution at 7-day lead times for RFEv2 and CHIRPS but retain it for IMERG. As JJAS is the main rainy season over Western Africa, this could arise from known challenges in forecasting West African monsoon rainfall, where convection is strongly organized by mesoscale convective systems that are poorly resolved in both the initial conditions and the physics of global NWP systems, degrading skill for the physical model in particular \citep{Vogel2018}. AI NWP models nonetheless show a strong skill improvement relative to IFS for this season, especially at longer lead times.

Out of all models and observational products, CHIRPS results for IFS show quite large uncertainty bands across all seasons. We note that this could arise from another known artefact of CHIRPS, where climatologically low-rainfall areas have an overestimation in number of rainfall events \citep{dinku2018}. Consequently, where the ``Only Wet" mask includes a greater share of marginal rainfall areas, we would expect noisier, less reliable verification against CHIRPS. Nonetheless, AI NWP models do not show such large uncertainties, and further analysis with more years of data would be warranted as to whether such are spurious artefacts indeed.

\begin{figure}[!ht]
 \centering
\includegraphics[width=\textwidth]{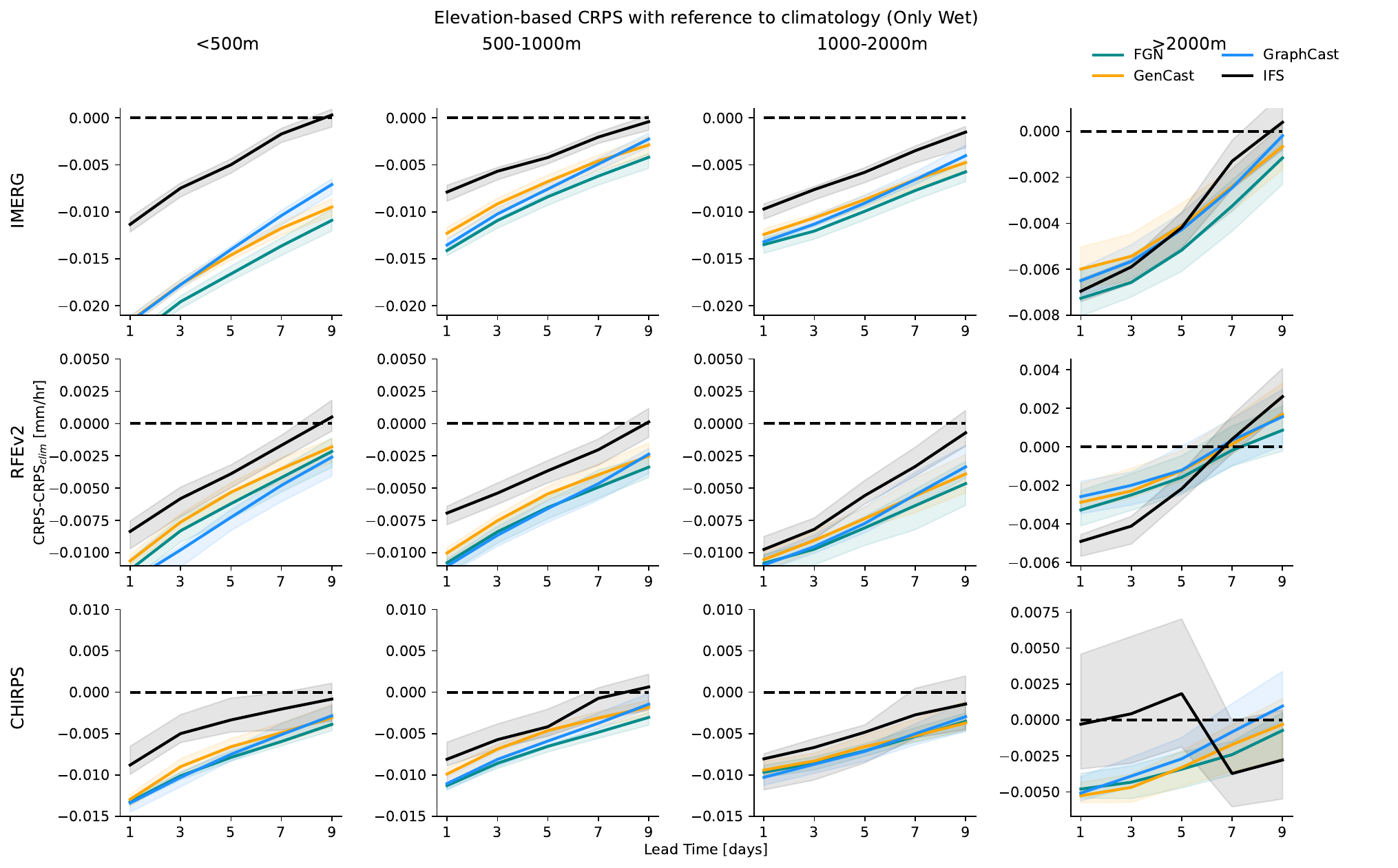}
 \caption{Same as Figure \ref{fig:fig3}, but showing elevation-wise performance when only considering areas receiving significant rainfall for each month instead. Elevation bins of $<$500~m, 500~m-1000~m, 1000~m-2000~m and $>$2000~m are considered.}
\label{fig:fig5}
\end{figure}

Figure \ref{fig:fig5} considers ``Only Wet" area $\Delta$CRPS$_{\mathrm{pot}}$ decomposed across elevation bins of $<$500~m, 500~m--1000~m, 1000~m--2000~m and $>$2000~m. For IMERG and CHIRPS, the gap between IFS and the AI NWPs narrows with increasing elevation, while RFEv2 shows a slight widening from $<$500~m to 500~m--1000~m before narrowing again. IMERG retains a notable improvement in skill,  beyond range of uncertainty bands, for the AI NWPs relative to IFS at 1000~m--2000~m, whereas RFEv2 and CHIRPS show less pronounced improvement here, with bands overlapping. At elevations above 2000~m, IMERG and RFEv2 show no meaningful difference in performance across models, with IFS even slightly outperforming the AI NWPs for RFEv2 at lead times under 7 days. For CHIRPS at $>$2000~m, IFS shows a sharp, transient improvement at 5--7 day lead times before returning to a near-climatological level by day 9, albeit with substantial uncertainty bands. Given the small number of high-elevation, ``Only Wet" grid cells in the $>$2000~m bin, this is plausibly due to the aforementioned CHIRPS artefact of low-rainfall areas having an overestimation in number of rainfall events, compounded by the small sample size available within the $>$2000~m bin. Distinguishing this from a genuine, lead-time-variant performance in IFS skill would however require further investigation.

\begin{figure}[!ht]
 \centering
\includegraphics[width=\textwidth]{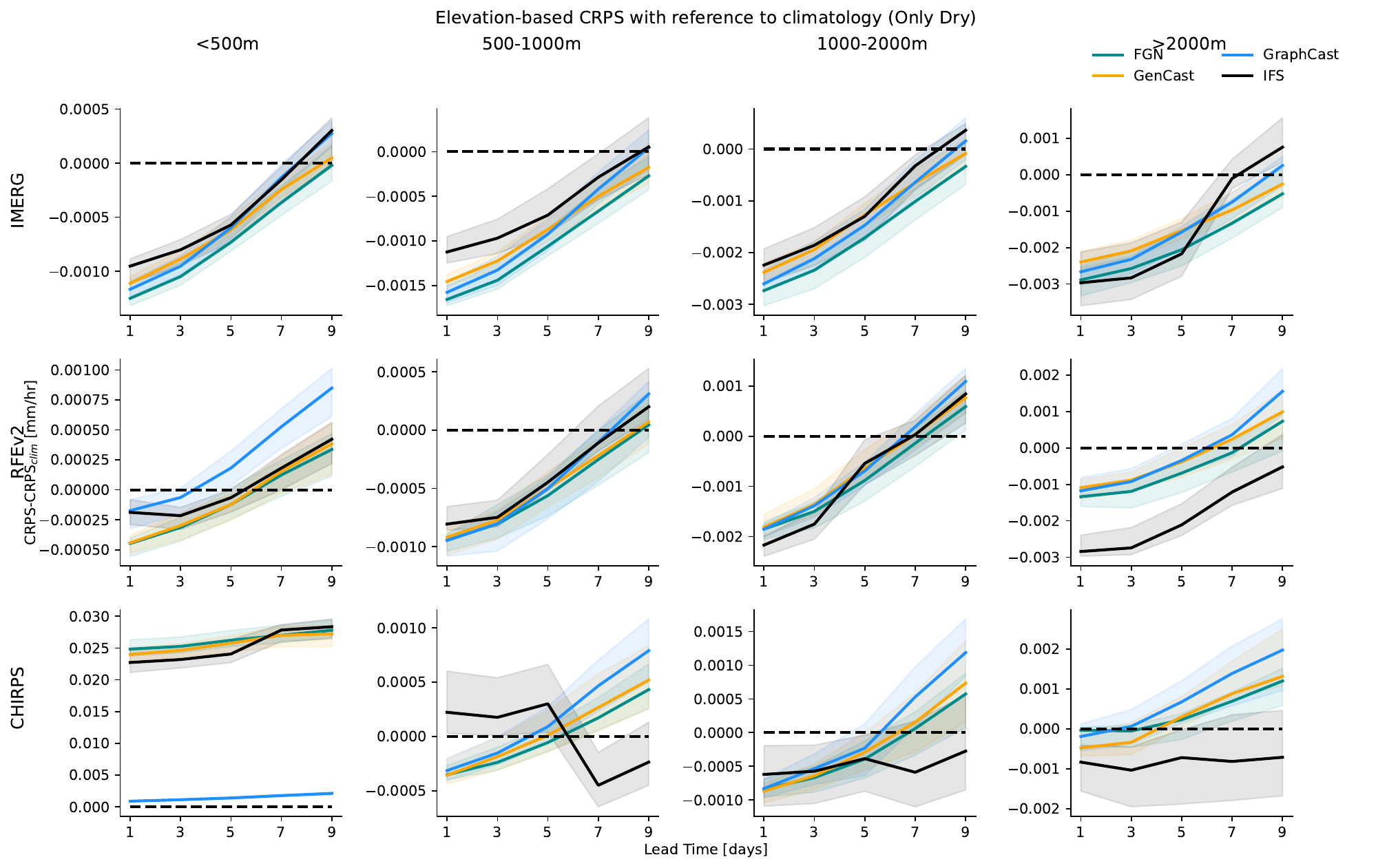}
 \caption{Same as Figure \ref{fig:fig5}, but now considering only dry areas.}
\label{fig:fig6}
\end{figure}

Generally, given that IFS runs on a 0.1$^{\circ}$ grid against the AI NWPs' 0.25$^{\circ}$, one might expect it to resolve finer scale processes that affect rainfall over higher, more complex terrain and have better performance. Our results of Figure \ref{fig:fig5} show a muted outcome here, which could be due to the IDR-recalibration closing the skill gap that would otherwise be expected due to IFS's finer model grid. To better disentangle this, we also consider the ``Only Dry" area $\Delta$CRPS$_{\mathrm{pot}}$ decomposed across elevation bins as shown in Figure \ref{fig:fig6}. Here, at 1000~m--2000~m and above, IFS shows a clear and consistent advantage over the AI NWPs for CHIRPS and RFEv2. This highlights that the physical drivers of dryness at higher elevations, such as orographic subsidence and rain-shadow effects are more explicitly resolved within IFS's native 0.1$^\circ$ grid resolution and therefore better represented in comparison to AI NWPs. In contrast, the convection responsible for most rainfall over ``Only Wet" areas remains a subgrid, parameterized process, making the gains when moving from the AI NWPs' 0.25$^\circ$ to IFS's 0.1$^\circ$ grid negligible. Similar to the skill jump seen at $>2000$~m for ``Only Wet" regions in Figure \ref{fig:fig5}, CHIRPS shows a sharp skill improvement at 5--7 day lead times for 500-1000~m in ``Only dry" regions. Rather than actual skill gain we attribute this to spurious artefacts arising from CHIRPS's known biases of clipping areas with climatological mean values of 0 to 0 \citep{dinku2018}, however further research with more years of data would help substantiate this.

\clearpage
\subsection{Spatial distribution and significance}\label{sec3b}

While $\Delta \mathrm{CRPS}_{\mathrm{pot}}$ provides insight into forecast resolution beyond climatology, assessing whether AI NWPs are fit for adoption requires detailed comparison of their relative skill gain against existing, more established systems. For this, we consider the CRPSS (Equation~\ref{eq:crpss-ifs}) using IFS as reference, with the overall IFS CRPS scores used in the denominator of the CRPSS calculation further provided in Figures \ref{fig:s3}-\ref{fig:s5}. As we have a continental focus, we choose IFS as the reference global physical NWP system, however note that national forecasting centres also employ the use of the Weather Research and Forecasting (WRF) system \citep{skamarock2019}. Given the different configurations of setup across the region and limited archival, we do not consider WRF here, however acknowledge it as a possible future point of reference for more bespoke country-level analysis, subject to sufficient volumes of archived data.

\begin{figure}[!ht]
 \centering
\includegraphics[width=\textwidth]{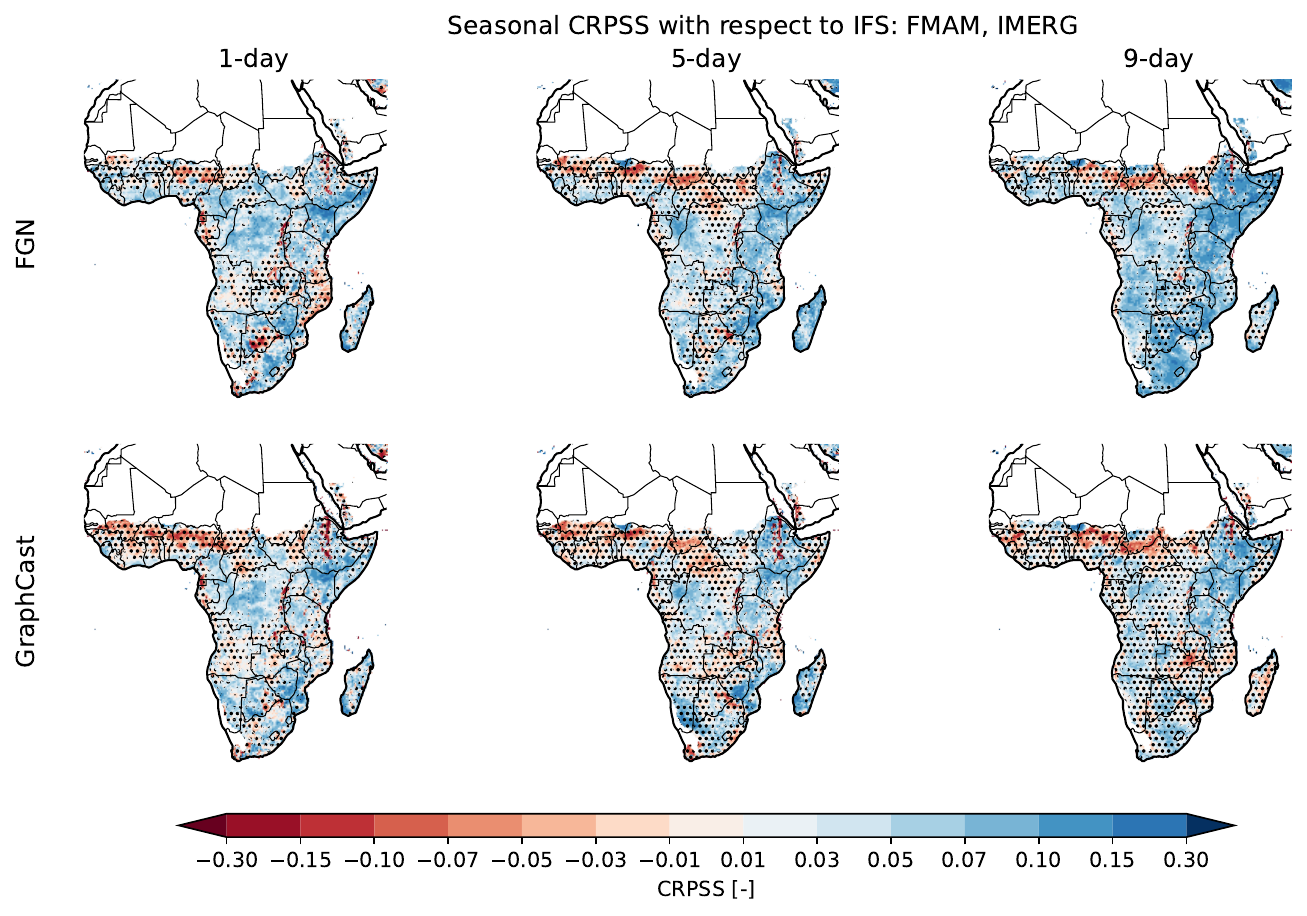}
 \caption{Spatial maps of potential Continuous Ranked Probability Skill Score (CRPSS) for FGN and GraphCast with reference to IFS and evaluated against IMERG for February March April and May (FMAM). 1- (top), 5- (middle) and 9- (bottom) day lead times are shown. A more positive CRPSS indicates greater skill improvement over IFS and vice versa for negative. Dotted areas show where the null hypothesis from the Diebold-Mariano test is not rejected and the skill differential therefore could be due to random chance.}
\label{fig:fig7}
\end{figure}

\begin{figure}[!ht]
 \centering
\includegraphics[width=\textwidth]{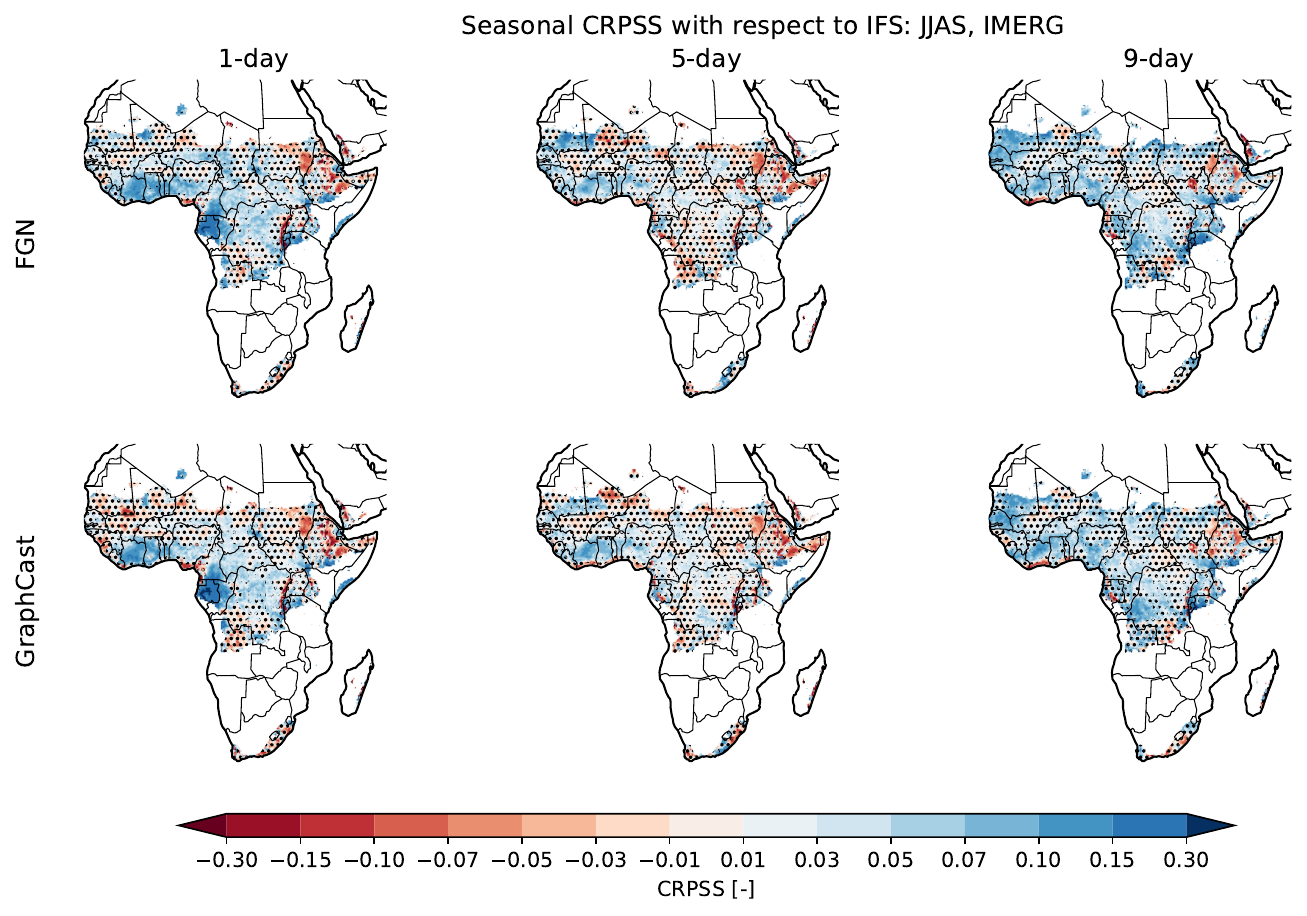}
 \caption{Same as Figure \ref{fig:fig7} but for June, July, August and September (JJAS).}
\label{fig:fig8}
\end{figure}

\begin{figure}[!ht]
 \centering
\includegraphics[width=\textwidth]{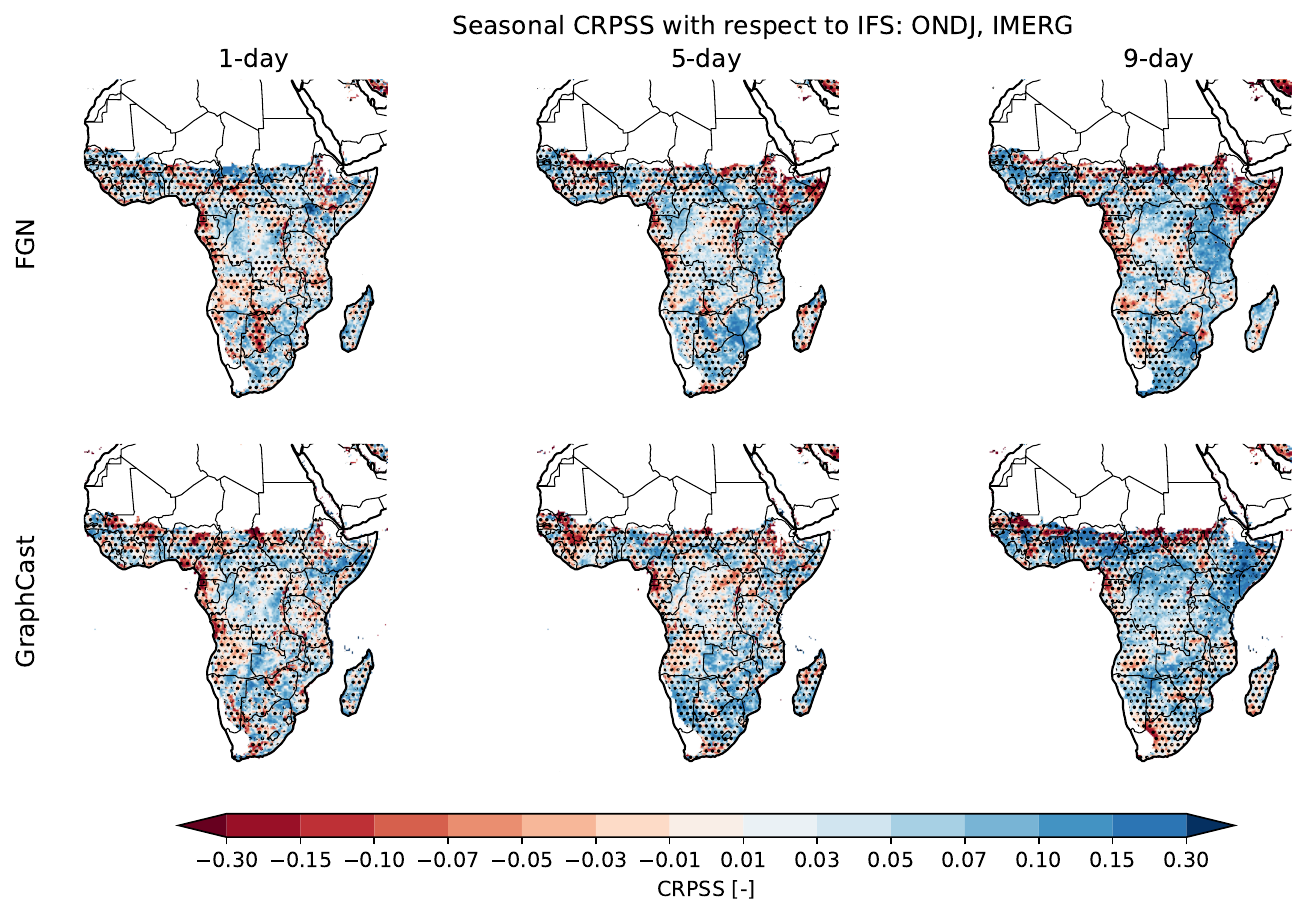}
 \caption{Same as Figure \ref{fig:fig7} but for October, November, December and January.}
\label{fig:fig9}
\end{figure}

Figures \ref{fig:fig7}-\ref{fig:fig9} provide maps of CRPSS values obtained for FGN and GraphCast when considering IMERG as ground truth, selected as the more optimistic product for assessing potential AI NWP skill gain given results of Section \ref{sec3a}. Maps for 1-, 5- and 9- day lead times across seasons of FMAM (Figure \ref{fig:fig7}), JJAS (Figure \ref{fig:fig8}) and ONDJ (Figure \ref{fig:fig9}) are shown with dry areas masked out. Dotted areas indicate where no statistically significant difference in skill was found (i.e., null hypothesis of the Diebold-Mariano test was not rejected). CRPSS values are calculated only for years 2024-2025, thus guaranteeing to be largely out of sample for both GraphCast and FGN. 

Across all lead times and seasons, both AI NWPs show promising skill gain over most of the continent. Compared to other regions, the Congo Basin shows the lowest but most consistent improvements in skill across seasons of around 3-5$\%$ which remain constant with lead time. This area is fundamentally limited by sparse observational networks affecting their initial condition generation \citep{Kammalac2026}, and while AI NWPs show promising improvements, they may also face an upper bound of improvement due to forecast errors associated with inaccurate initial conditions. For both FMAM and ONDJ, highest skill gains occur in the Eastern and Southern parts of Africa, with values of 7-10$\%$ improvement. This improvement is more striking with greater areas of statistical significance in FMAM than ONDJ. FMAM is known to be less predictable than ONDJ \citep{Camberlin2002,Nicholson2014,Nicholson2015} and these results provide an encouraging signal of improvement, although further investigation will be needed dissect the exact reasons. JJAS shows most improvement over Western Africa, with a notable skill gain of $>15\%$ around Cameroon and Gabon at 1-day lead times. This skill gain however rapidly degrades to 2-5~$\%$ with increasing lead times, which could be attributed to the general unpredictability of the organised mesoscale convective systems driving rainfall within the region. Combined with poor initial conditions, which are limited by sparse observations such error growth is more rapid with lead time \citep{Vogel2018,Kammalac2026}.

Slight degradations in performance can be noted along the Sahel belt for FMAM and ONDJ, as well as along the Ethiopian Highlands in JJAS. Nonetheless, most areas showing degradation are indicated as not statistically significant, and this could therefore instead be due to sampling variability. For FMAM and ONDJ, areas with no statistically significant difference in skill increases with lead time, which is consistent with the narrower skill gap between AI NWPs and IFS with increasing lead time seen for IMERG in Figure \ref{fig:fig4}. This increase is more pronounced for GraphCast than for FGN, potentially because averaging over FGN's ensemble members dampens the growth of error with lead time, whereas GraphCast's single deterministic trajectory is unconstrained by ensemble averaging, causing its skill to degrade quicker. Interestingly the Eastern African coastline, spanning Ethiopia down to Tanzania retains a significant skill gain across all lead times for both FGN and GraphCast. This can be attributed to the known predictability over the region, with ONDJ displaying high correlations between rainfall and equatorial Indian Ocean westerlies \citep{hastenrath1993,hastenrath2004}, while correlation for rainfall patterns in FMAM have been found, albeit not as consistent \citep{Camberlin2002,Nicholson2015}.

\begin{figure}[!ht]
 \centering
\includegraphics[width=\textwidth]{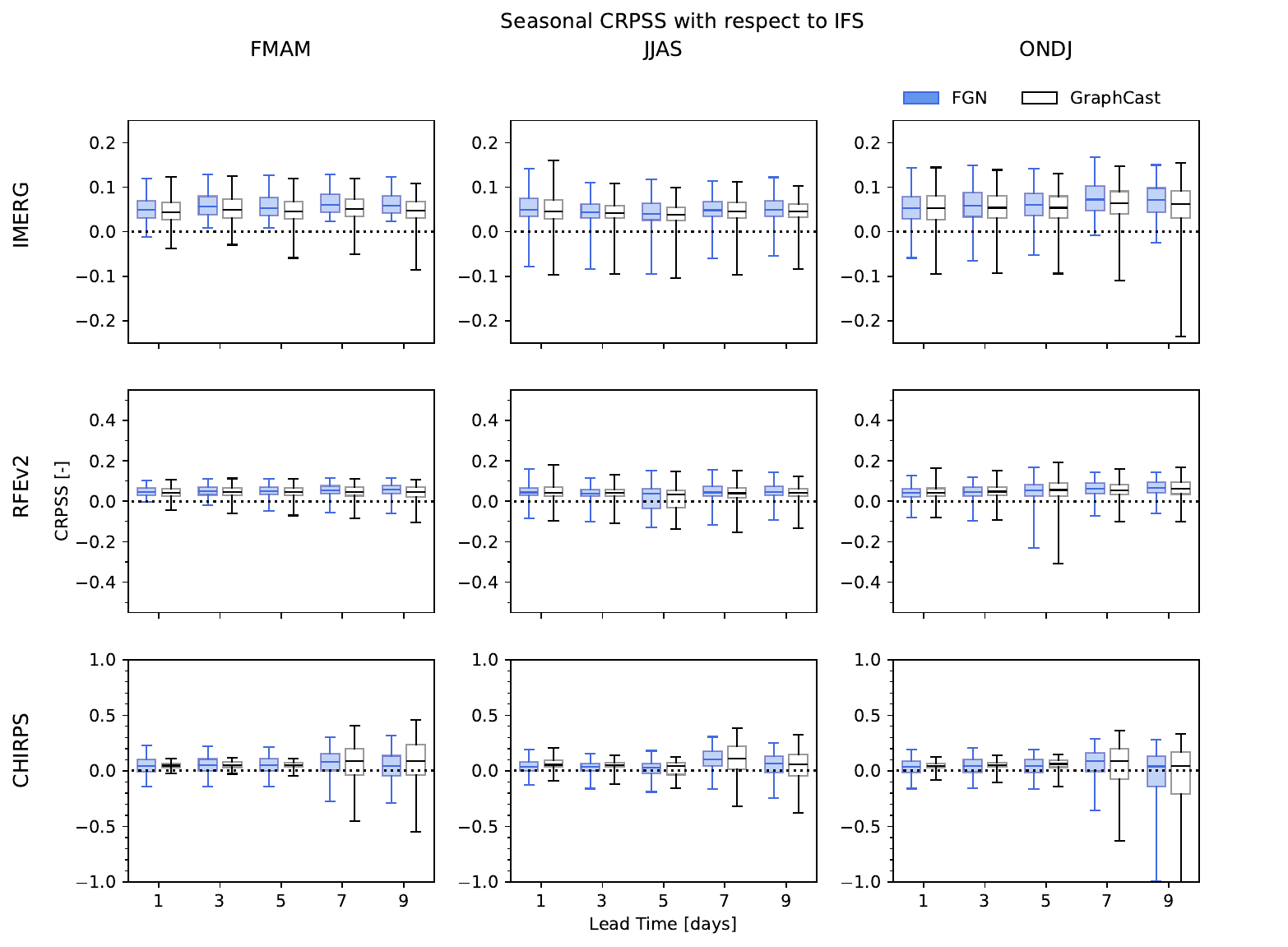}
 \caption{Box-plots of the potential Continuous Ranked Probability Skill Score (CRPSS) for FGN and GraphCast with reference to IFS evaluated across 1-, 3-, 5-, 7- and 9- day lead times. Seasonal performance for February March April and May (FMAM, left); June, July, August and September (JJAS, middle); and October, November, December and January (ONDJ, right) when considering observations of IMERG (top), RFEv2 (middle) and CHIRPS (bottom) are shown. Only grid-cells with a significant loss differential as obtained by the Diebold-Mariano test are considered. Whiskers indicate 90$\%$ spatial spread across grid cells.}
\label{fig:fig10}
\end{figure}

\begin{figure}[!ht]
 \centering
\includegraphics[width=\textwidth]{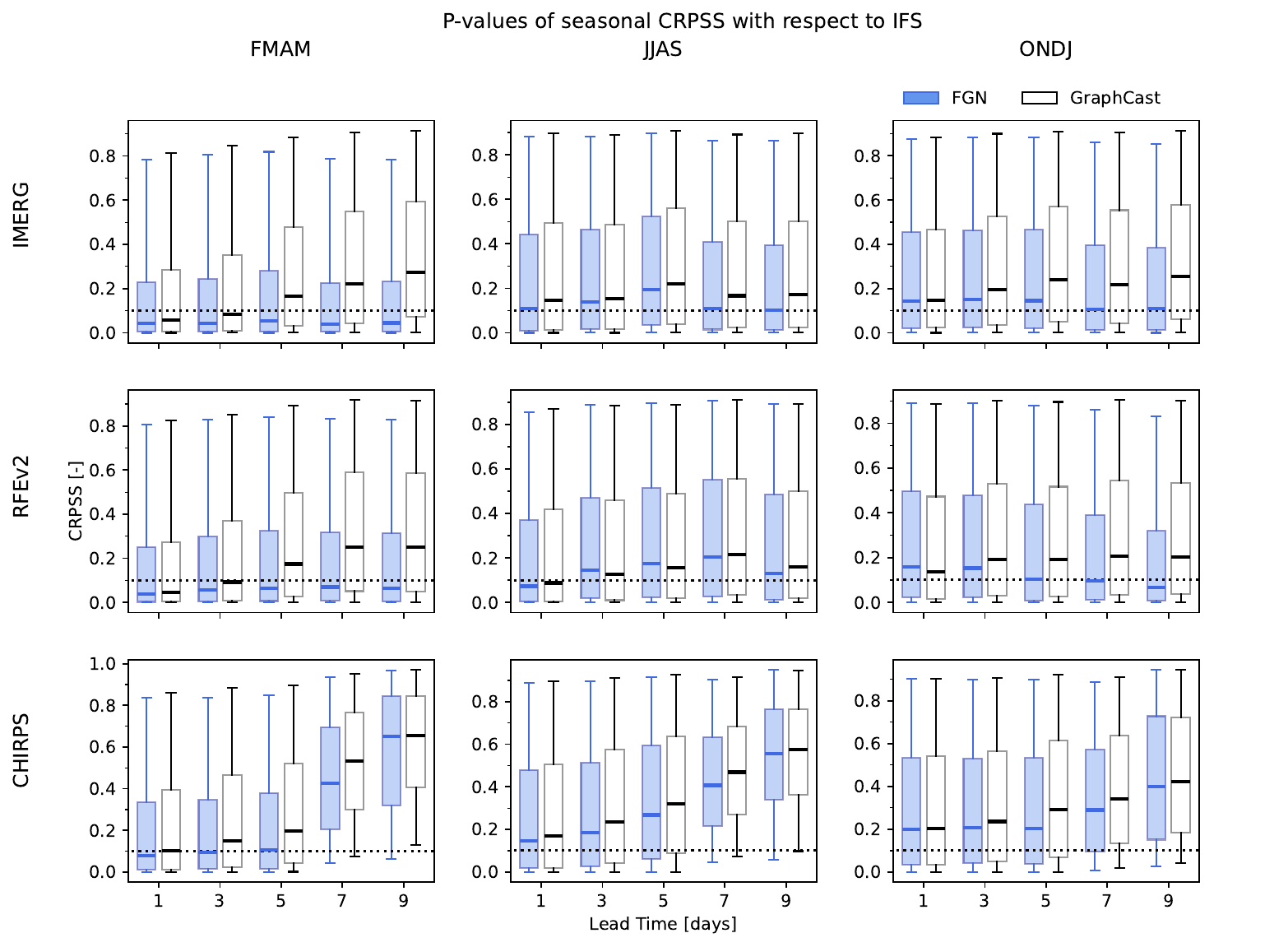}
 \caption{Box-plots of p-values obtained from the Diebold-Mariano test applied to FGN and GraphCast with reference to IFS evaluated across 1-, 3-, 5-, 7- and 9- day lead times. Seasonal performance for February March April and May (FMAM, left); June, July, August and September (JJAS, middle); and October, November, December and January (ONDJ, right) when considering observations of IMERG (top), RFEv2 (middle) and CHIRPS (bottom) are shown. P-values lower than 0.1 (dotted horizontal line) indicates that the potential CRPS loss differential to IFS is significant. Whiskers indicate 90$\%$ confidence intervals.}
\label{fig:fig11}
\end{figure}

Figures \ref{fig:fig10} and \ref{fig:fig11} provide box-plot summaries of seasonal CRPSS values and their corresponding $p$-values respectively, obtained at a grid-cell level over the region for all observational products, across 1-, 3-, 5-, 7- and 9- day lead times. Note that the CRPSS boxplots in Figure \ref{fig:fig8} are only plotted for areas with a statistically significant change in skill with a significance level of 10$\%$ used (i.e., $p$-values less than 0.1). Both FGN and GraphCast display median CRPSS values close to 0.05 (i.e., $\sim{5}\%$ skill gain relative to IFS) across all lead times and observational products, with the exception of CHIRPS at 7- and 9- day lead times for FMAM and JJAS. For FMAM and JJAS, both IMERG and RFEv2 show similar inter-quartile ranges, roughly between 0.025 and 0.075 and well above 0, with the exception of RFEv2 at 5- (7- and 9-) day lead times in JJAS (FMAM). ONDJ generally shows larger performance gains for both IMERG and RFEv2, with the upper quartiles reaching 0.1 and lower quartiles being above 0 across all lead times, albeit larger 90$\%$ confidence intervals, although this should be weighed against the lower amount of statistically significant areas found for this region as seen in Figure~\ref{fig:fig9}. $p$-values for both IMERG and RFE show most significant skill improvement for FGN as compared to GraphCast. IMERG in particular shows more than 50$\%$ of grid-cells with statistically significant changes in skill ($p$-values less than 0.1) across all seasons and lead times, with the exception of 3- and 5- (1- to 5-) day lead times for JJAS (ONDJ). FMAM shows the most strikingly significant changes in skill for IMERG, FGN corroborating the skill improvement maps from Figure \ref{fig:fig7}. In comparison the improvement observed for RFEv2 is more modest with at least 25$\%$ if not more grid-cells having statistically significant changes in skill across seasons and lead times.

CHIRPS shows more varied performance compared to the other products, with FMAM and ONDJ displaying lower quartiles of at least -0.25, indicating areas of skill degradation, while upper quartiles still indicate notable skill improvement with values up to 0.3. JJAS shows a more consistent performance gain over IFS with lower quartiles mostly above 0 across all lead times and upper quartiles reaching up to 0.2. $p$-values obtained for CHIRPS show lower statistical significance relative to IMERG and RFEv2, with around 25$\%$ of grid-cells having statistically significant changes in skill for 1- to 5- day lead times across all seasons, which then degrades to less than 25$\%$ at 7- and 9- day lead times. This degradation is most emphasised for FMAM and JJAS and could arise from aforementioned artefacts in CHIRPS, that lead to lower reliability in verification results for low-rainfall areas. That this disproportionately affects longer lead times could suggest that as the forecast skill degrades with lead time, the effectiveness of re-calibration under noisy target data also degrades more rapidly, however further investigation would be required here.

\section{Conclusion and Discussion}\label{sec4}
This study seeks to provide a fair assessment of the potential operational utility of AI NWP rainfall forecasts considering the models, GraphCast, GenCast and FGN. Such an exercise requires imitation of what an actual forecasts' quality would be when reaching an operational forecaster's desk. In such, any AI or physical NWP output would first have to be post-processed so as to match the resolution most relevant to a given national meteorological centre as well as be corrected towards suitable, local observational products. We employ an Isotonic Distributional Regression (IDR)-based approach to post-process both the physical NWP, IFS, and AI NWP rainfall forecasts to trusted, observational products over Africa, namely, IMERG, RFEv2 and CHIRPS. Beyond simple downscaling and post-processing of NWP outputs, the IDR provides an optimal solution to probabilistic scoring metrics such as the Continuous Ranked Probability Score (CRPS) \citep{Henzi2021}, and outputs calibrated cumulative distribution functions that can be used to estimate a forecast's potential calibrated skill regardless of whether the initial forecast model was deterministic or probabilistic. This further allows for estimation of forecast miscalibration and resolution following Brier-skill score decompositions generalised to proper scoring rules \citep{Siegert2017, Dimitriadis2021}.

When considering the forecast resolution, estimated as the IDR-calibrated performance against that of climatology, all NWPs show skill beyond climatology across seasons. JJAS proves to be the season most difficult to forecast with IFS losing resolution across observational products at 5-7 day lead times, while AI NWP models retain resolution at longer lead times only degrading beyond climatology at 7-9 days. We note that making a distinction between calculating the skill over areas receiving significant rainfall (simply called ``Only Wet", with the opposite being ``Only Dry") offers a more consistent view across observational products and this could arise from known observational product-dependent artefacts such that low-rainfall areas can display noisy, unreliable results. In further assessing the elevation-dependent performance, we expected IFS to perform better for higher, more complex terrain due to its finer native model grid. While we note no significant gain from IFS at higher elevations for ``Only Wet" areas, both CHIRPS and RFE2 show improvements from IFS at $>$2000~m within ``Only Dry" areas. This suggests that processes governing dryness such as subsidence are resolved within the 9~km resolution of IFS with notable benefits therefore over the AI NWP models. On the other hand, typical scales for convection (around 4~km) that drive rainfall are still within the sub-grid level of IFS and hence no obvious benefit over AI NWPs is seen there.

One striking result when comparing the overall skill within AI NWPs is that across seasons and elevations, the calibrated performance of GraphCast is almost on par with that of FGN. Moreover when considering spatial CRPSS values for areas with statistically significant changes in skill, both FGN and GraphCast show similar median improvements of around 5$\%$ across all observational products. This could suggest that probabilistic correction provides just as much improvement when applied to a deterministic versus a probabilistic model backbone. However, when considering amount of area with statistically significant changes in skill, a greater significance in skill gained for FGN at longer lead times is observed as compared to GraphCast. This emphasises a key advantage of ensemble systems \citep{Mahesh2025}, where error growth, when considering the ensemble average, is dampened as compared to a single deterministic trajectory, whose errors compound more rapidly with lead time. An ensemble is not only a device for reducing error in the mean, however: many downstream applications ingest forecasts as fields or trajectories rather than as pointwise distributions, and depend on members that remain coherent in space and time. Calibration applied to a summary statistic recovers the marginal distribution at each grid point but not that joint structure, so the equivalence reported here is an equivalence in pointwise calibrated skill rather than in the information a user can act on. 

This study's results should be considered with some limitations. Firstly, by concentrating on the overall calibrated performance, the potential skill differences at extreme rainfall thresholds may be hidden. Given the threshold-calibrated property of IDR \citep{Henzi2021}, the skill gains are expected to hold even at higher rainfall thresholds \citep{nath2026predicting}. Nevertheless, across literature, the performance of AI- against physical- NWP for extremes has mixed conclusions \citep{Shi2025,Zhang2026} and a dedicated further study using threshold -weighted as well as -specialised skill scores would be warranted. Moreover, when considering rainfall where double penalty problems may arise \citep{Hagelin2017,Jain2023}, spatial verification metrics should be further considered \citep{Guilpart2025,loveday2026,Wang2026} although, these are not regarded as ``proper" scores and would require further consideration when applying score decompositions. The second limitation is inherent to the present evaluation period. Data availability constrains the years available for verification unevenly across models: IFS, 2023 was left out due to retrieval overheads, while FGN additionally lacks 2020 and 2021. A further concern is that of data leakage, as the AI NWP models considered here were trained on and fine-tuned on data whose period may overlap with that used for evaluation, potentially inflating their apparent skill. Nonetheless, when considering the skill relative to the physical model, IFS – which is not subject to this risk in the same way – we only considered years 2024-2025 which are largely out of sample and still provided consistent performance gains. However, as AI NWP models continue to be trained on ever larger and more recent stretches of the observational record, the pool of years remaining entirely out-of-sample for fair evaluation will only continue to reduce. Addressing this is likely to require closer coordination between forecast providers and modellers on held-out evaluation periods, or a broader shift toward prospective, real-time verification as the primary standard against which operational suitability is judged.

\section{Outlook}\label{sec5}

As data-driven models, AI NWPs need to be trained on as much data as available, in order to achieve their best possible performance. A natural step for modelling centres is therefore to continuously update newer training cycles of AI NWPs with the most recent observational records. Nevertheless, as the first line of dissemination, National Meteorological and Hydrological Services (NMHSs) require vetted, well-verified forecasts so as to maintain institutional accountability \citep{Pappenberger2015,Cloke2009}. More than ever, the new generation of AI NWPs, therefore warrants a closer coordination between forecast providers and modellers to conduct fair comparisons and quality control newer models as they come out. 

Practically speaking, a broader shift towards prospective, real-time verification as the primary standard against which operational readiness is judged may become a future staple to cope with the rapid AI NWP development and update cycles. Embedding such real-time verification systems within operational centres would necessitate verification to be treated not as a one-off academic exercise but as a continuous ``Quality Monitoring and Management System". NMHSs increasingly recognise routine, automated verification, paired with structured feedback channels between forecasters and model developers, as core to maintaining forecast quality over time \citep{Pagano2025}. Such feedback loops would need to explicitly track and flag when a model's training period begins to encroach on its evaluation period, rather than assuming this boundary remains fixed. Additionally, physical NWP systems such as IFS are themselves periodically upgraded, and such upgrades can alter the operational analysis used both to initialise forecasts and, for AI NWP models fine-tuned on this analysis, to define their training distribution. Ben Bouall\`egue et al. (2026) show that AI models incorporating a fine-tuning step, are measurably more sensitive to a change in IFS cycle, with performance degrading under a new analysis despite no change to the model itself \citep{BenBouallegue2026}. This necessitates a tighter coupling between forecast providers, model developers and operational forecasters than has traditionally been the case, so that emerging discrepancies between apparent and genuine skill can be identified and communicated as they arise, particularly as both physical and AI NWP systems continue to evolve \citep{Nath2026}. 

Throughout this study, we note systematic issues that may affect the quality and predictability of both physical and AI NWP forecasts at longer prediction horizons. While these can be linked to the known chaotic nature of the local as well as mesoscale, organised convective processes driving rainfall, they also elicit a few minimal actions that could be taken to achieve improvement in forecast quality, particularly at longer prediction horizons. The first, most obvious step is improvement in observational networks. Without a well-developed observational network, initial conditions over the continent suffer leading to rapid error growth within forecasts. While this could potentially be statistically ameliorated or dampened within AI NWP models, it is difficult to ascertain whether this is through learning truly, physically consistent corrections or by chance. Following this, a second step is to improve our understanding of the success and failure modes of AI NWPs. To this end, understanding the internal representations learned by these models would answer why and where they outperform traditional physics-based NWPs. This could create a tighter feedback loop between physical and AI NWP developments, as insights into the learned dynamics that drive AI skill gains could guide targeted improvements in physical NWPs. These enhanced physical NWPs would then yield better initial conditions and higher-quality training data for the next generation of AI NWPs.

%
%

\paragraph{Acknowledgements} We would like to thank Guy Shalev, Ilan Price, Stephan Rasp, Elinor Kruse and Tufa Dinku for their valuable feedback throughout the process of designing this study. Shruti Nath would like to thank Myles Allen and Antje Weisheimer for their continuous support.

\paragraph{Funding} This work was supported by Google.org through the Forecast4Africa programme, delivered by AfriClimate AI in partnership with National Meteorological and Hydrological Services across Africa. The funder had no role in the study design, data analysis, interpretation of results, or the decision to submit the manuscript for publication.

\paragraph{Author Contributions} Data acquisition and post-processing was done by DS and KA. SN designed the study and application of the calibration with support from DS and KA. SN conducted the analysis and write-up. JB, JK, RM, FP and FC provided further review and streamlining fo the text.

\paragraph{Data}All observational datasets used in this study are openly available. GPM IMERG Final Run V07 is distributed by the NASA Goddard Earth Sciences Data and Information Services Center (GES DISC), \url{https://disc.gsfc.nasa.gov}, with the daily product archived under doi:10.5067/GPM/IMERGDF/DAY/07. The IMERG data were provided by the NASA/Goddard Space Flight Center and the Precipitation Processing System, which develop and compute IMERG as a contribution to GPM, and archived at the NASA GES DISC. CHIRPS version 2.0 is produced by the Climate Hazards Center at the University of California, Santa Barbara, \url{https://www.chc.ucsb.edu/data/chirps}, and is placed in the public domain under a Creative Commons CC0 waiver; users are asked to cite Funk et al. (2015), doi:10.1038/sdata.2015.66. RFEv2 is produced by the NOAA Climate Prediction Center in support of the USAID Famine Early Warning Systems Network and is available at \url{https://ftp.cpc.ncep.noaa.gov/fews/fewsdata/africa/rfe2/}; as a work of the United States Government it carries no copyright, and the algorithm, together with the closely related ARC2 product, is documented by Novella and Thiaw (2013), doi:10.1175/JAMC-D-11-0238.1. The Köppen-Geiger climate classification used for the zonal pooling in Section~\ref{sec2c2} is that of Beck et al. (2018), doi:10.1038/sdata.2018.214, available under CC BY 4.0. IFS ensemble total precipitation fields were obtained from the WeatherBench2 archive, \url{gs://weatherbench2/datasets} (Rasp et al. 2024, J. Adv. Model. Earth Syst. 16, e2023MS004019), and from the ECMWF Meteorological Archival and Retrieval System (MARS). ECMWF real-time forecast products, including the IFS ensemble, are additionally available openly through ECMWF Open Data at \url{https://data.ecmwf.int/forecasts/} under the Creative Commons Attribution 4.0 International (CC BY 4.0) licence, copyright European Centre for Medium-Range Weather Forecasts (ECMWF), source \url{https://www.ecmwf.int}. ECMWF does not accept any liability whatsoever for any error or omission in the data, their availability, or for any loss or damage arising from their use. GraphCast, GenCast and FGN were developed by Google DeepMind and are described in the references cited in Section 2.1; code and model weights are distributed at \url{https://github.com/google-deepmind/weathernext}, under the Apache 2.0 licence for code and CC BY 4.0 for model weights.

\section*{Supplementary Data}
\setcounter{figure}{0}
\renewcommand{\figurename}{Fig.}
\renewcommand{\thefigure}{S\arabic{figure}}

\begin{figure}[ht!]
    \centering
    \includegraphics[width=\textwidth]{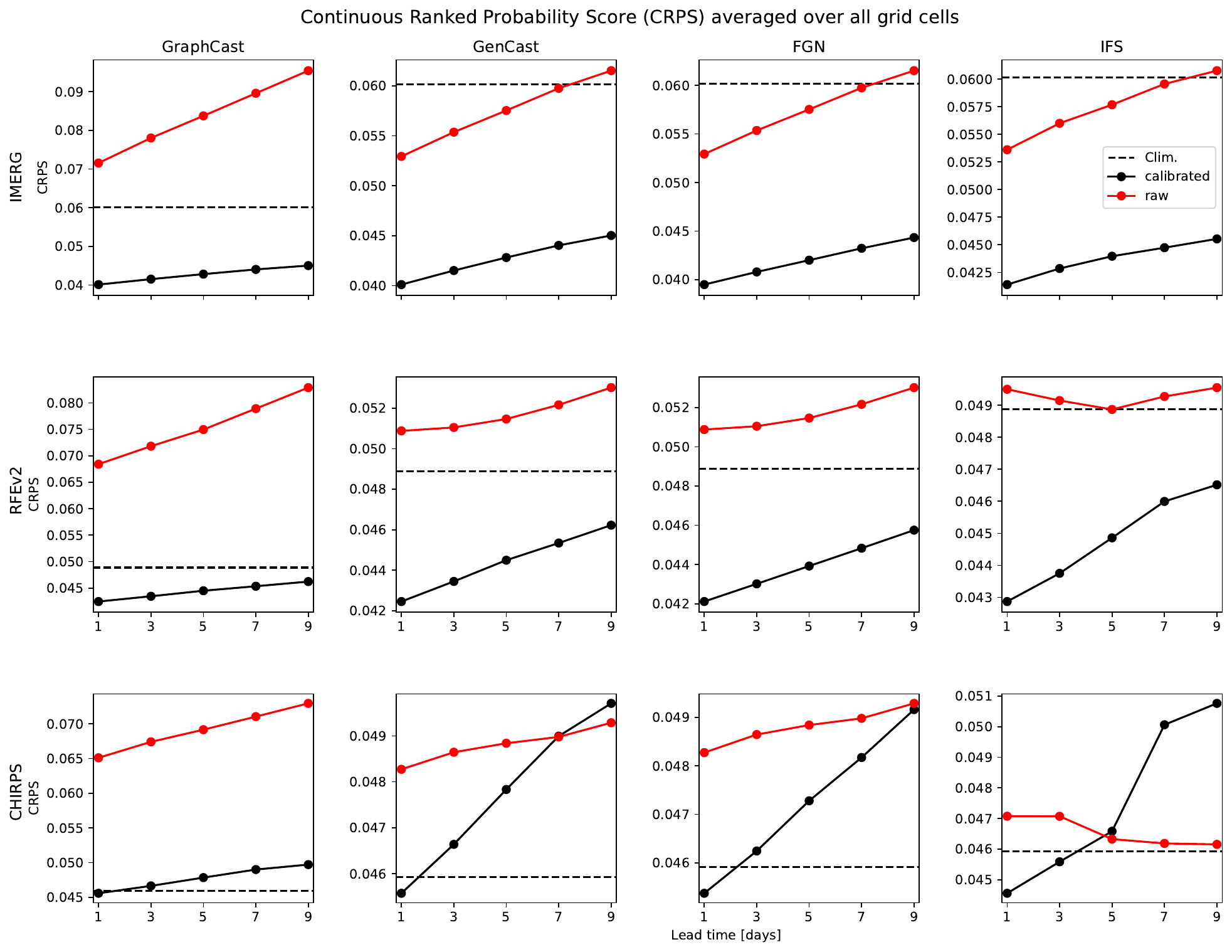}
    \caption{CRPS vs lead time for raw (red) and Isotonic Distribution Regression based (black) outputs of GraphCast, GenCast, FGN and IFS. CRPS is averaged over all grid points, climatology is provided as the dashed line.}
    \label{fig:s1}
\end{figure}

\begin{figure}[ht!]
    \centering
    \includegraphics[width=\textwidth]{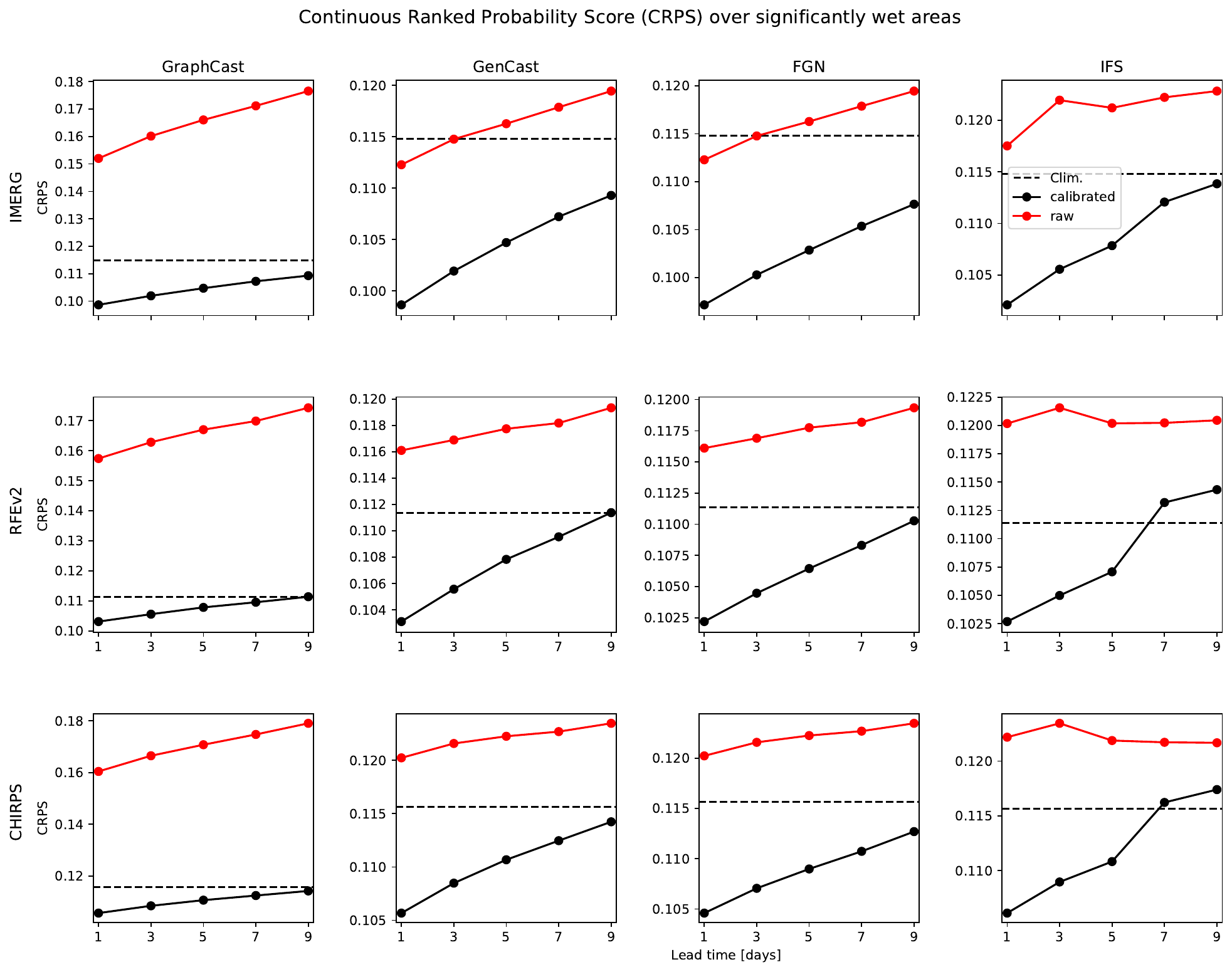}
    \caption{Same as Figure \ref{fig:s1}, but CRPS is averaged over only significantly wet areas for a given month, before then taking the annual average.}
    \label{fig:s2}
\end{figure}

\begin{figure}[ht!]
    \centering
    \includegraphics[width=\textwidth]{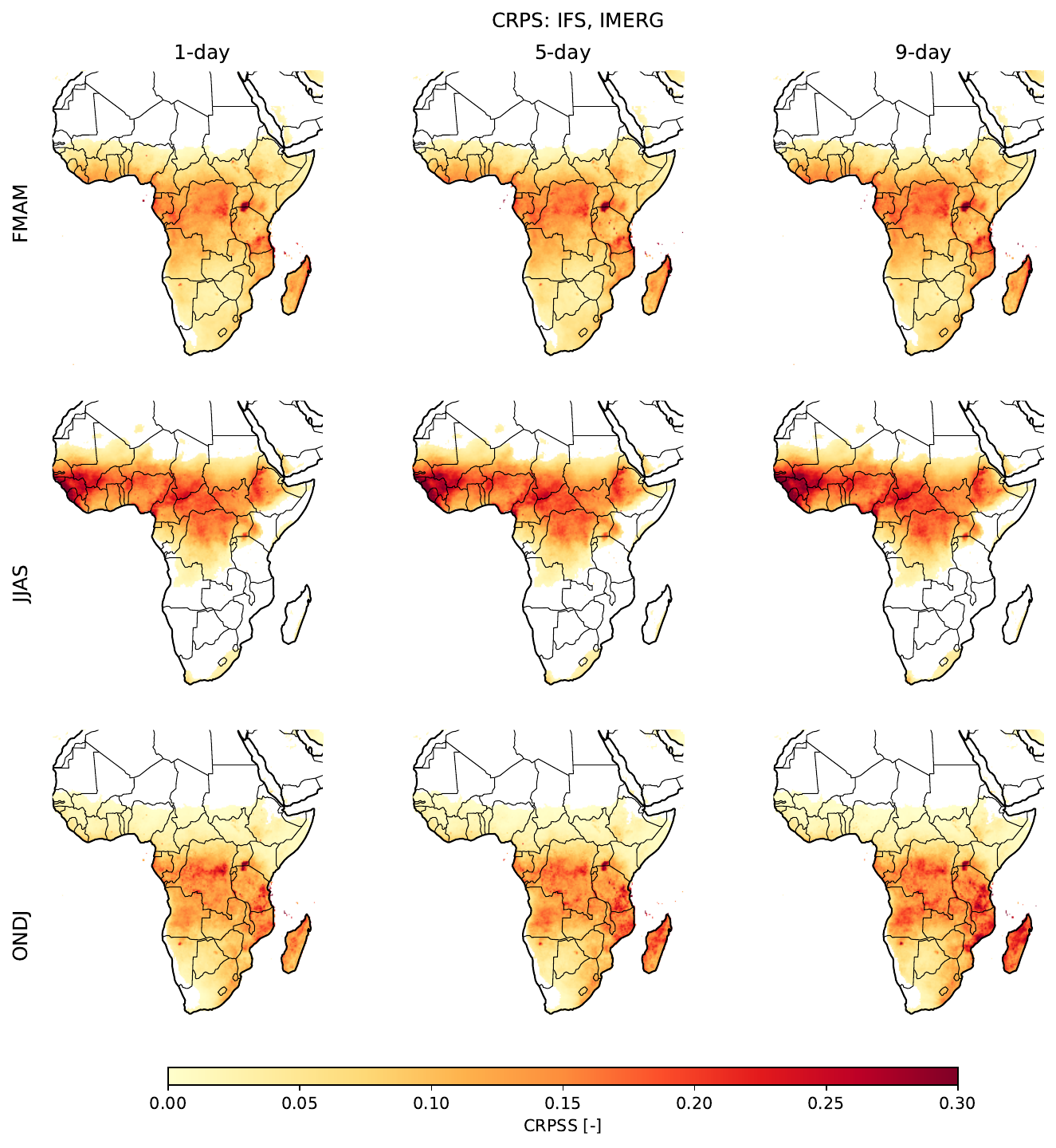}
    \caption{CRPS maps of IFS calibrated with Isotonic Distributional Regression towards IMERG. Lead times of 1-, 5- and 9- day lead (columns) and seasons of February, March, April and May (FMAM), June, July, August and September (JJAS) and October, November, December and January (ONDJ) (rows) are considered.}
    \label{fig:s3}
\end{figure}

\begin{figure}
    \centering
    \includegraphics[width=\textwidth]{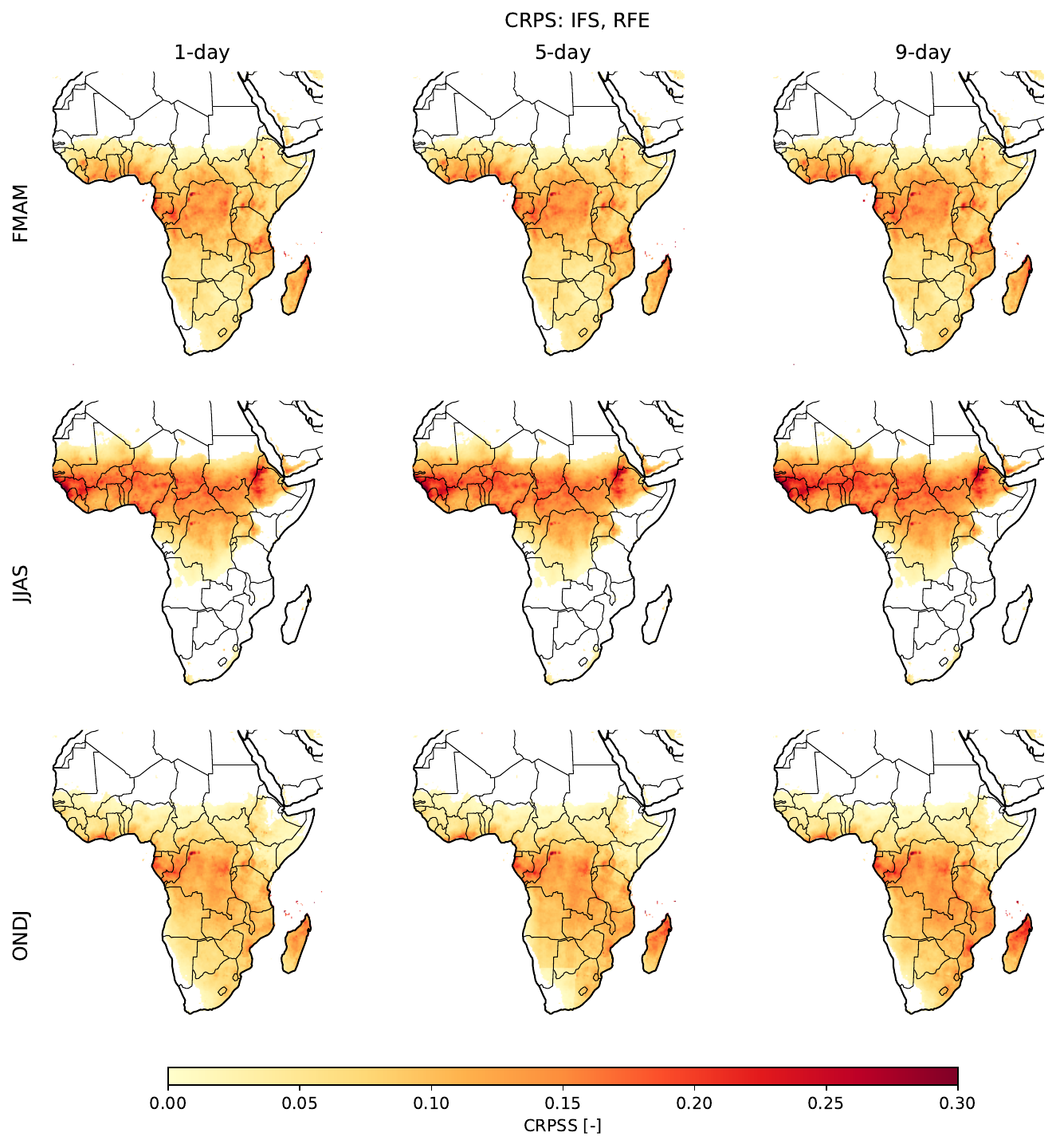}
    \caption{Same as Figure \ref{fig:s3} but for RFEv2}
    \label{fig:s4}
\end{figure}

\begin{figure}
    \centering
    \includegraphics[width=\textwidth]{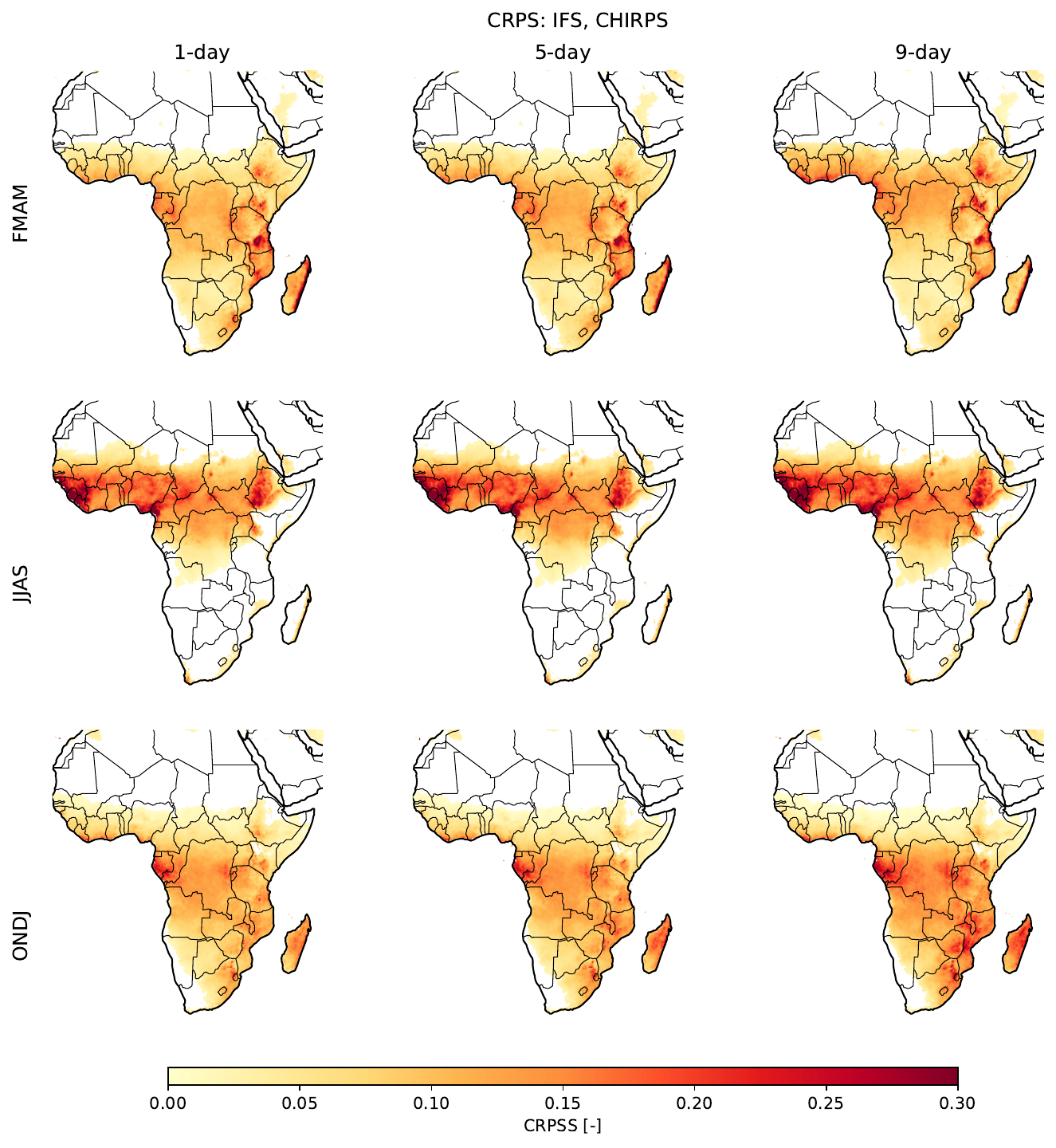}
    \caption{Same as Figure \ref{fig:s3} but for CHIRPS}
    \label{fig:s5}
\end{figure}
\clearpage

\bibliography{references}
\bibliographystyle{plainnat}

\end{document}